\documentclass[11pt,a4paper]{scrartcl}
  \usepackage[greek,pdf]{preamble}
  \usepackage{abstract}
  \title{\vspace{-2cm}Engines of Parsimony: Part~II}
  \subtitle{Performance Trade-offs for Communicating Reversible Computers}
\endofdump

  \DeclareMathOperator*{\medotimes}{\text{\raisebox{0.25ex}{\scalebox{0.8}{$\bigotimes$}}}}
  \usepackage{centernot}
  \usepackage{bbold}
  \DeclareSIUnit\molar{\mole\per\cubic\deci\metre}
  \DeclareSIUnit\Molar{\textsc{m}}
  \def\rate{\operatorname\lambda}

  \def\largertitlepage{\thispagestyle{empty}\enlargethispage{2\baselineskip}}

\begin{document}
\maketitle
\largertitlepage

\begin{figure}[h!]
  \vspace{-1.5cm}
  \centering
  \includegraphics[width=.4\textwidth]{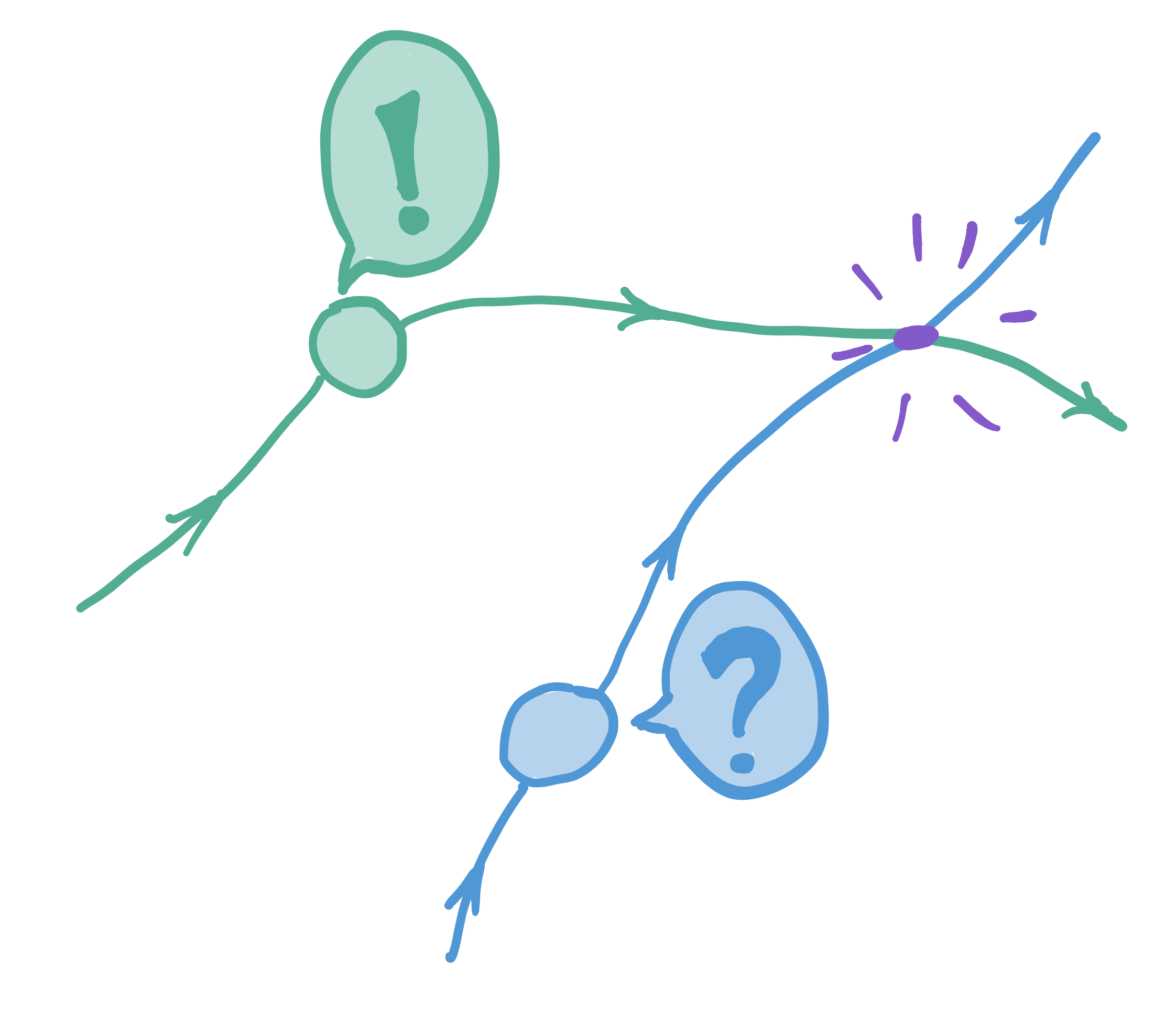}%
  \includegraphics[width=.4\textwidth]{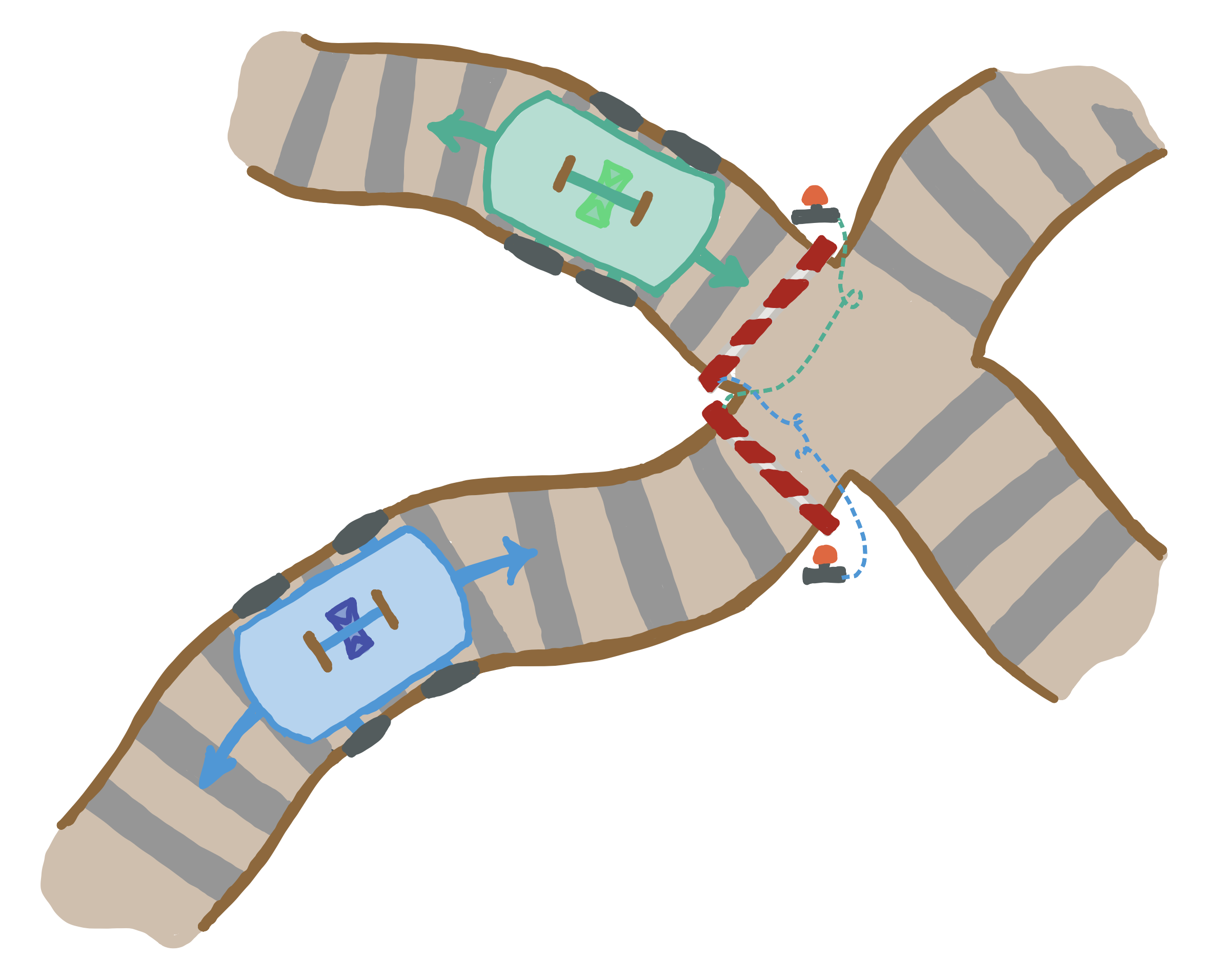}
    
  \caption{This paper concerns synchronisation interactions between reversible Brownian computational particles, which can walk backwards and forwards through phase space. This is illustrated here as two wagons on intersecting train tracks, where there is a barrier at the intersection. Each wagon controls the other's barrier, and so both must be present simultaneously in order to pass through the intersection.}
  \label{fig:cover}
\end{figure}

\renewcommand{\abstractname}{Lay Summary}
\begin{abstract}
  In recent years, unconventional forms of computing ranging from molecular computers made out of DNA to quantum computers have started to be realised. Not only that, but they are becoming increasingly sophisticated and have a lot of potential to influence the future of computing. One interesting class of unconventional computers is that of reversible computers, which includes quantum computers. Reversible computing---wherein state transitions must be invertible, and therefore must conserve information---is largely neglected outside of quantum computing, but is a promising avenue for realising substantial gains in computational performance and energy efficiency. 

  In Part~I~\cite{earley-parsimony-i} of this series on performance of reversible computers, a robust scaling law was found describing the maximum performance of a reversible computer---in terms of operations per second---that can be extracted from a given region of space supplied with a given input power and heat dissipation flux, taking into account classical, quantum and relativistic physics. The scaling law was found to be $\sim\sqrt{PM}$ where $P$ is the heat dissipation rate (and should balance the input power) and $M$ is the enclosed mass, confirming and strengthening a result of \textcite{frank-thesis}, whilst the equivalent law for irreversible computers is $\sim P$ due to the Landauer bound~\cite{landauer-limit,szilard-engine}. In practice this manifests as $\sim V^{5/6}$ and $\sim V^{2/3}$ respectively as $P\sim A\sim V^{2/3}$ where $A$ is the convex bounding surface area of the region and $V$ its volume, and so reversible computers significantly outperform irreversible computers.

  This analysis neglected, however, to consider interactions between subunits of these computers. In particular, a large computer will be composed of many small computational subunits. Each of these subunits will perform independent computation, but for useful programs one often wishes to combine the results of these independent computations. To do so, the subunits must communicate---they must synchronise their individual states. In an irreversible computer with ample free energy density, this is a non-issue. In contrast, when free energy is limiting, the hidden entropic cost of synchronisation is made manifest and must be considered.

  To illustrate, consider two reversible Brownian computers which would like to communicate with one another. A reversible Brownian computer is free to wander back and forth along its configuration space, the set of possible computational states it can take, and we can illustrate this by the analogy of two wagons on intersecting train tracks as shown in \Cref{fig:cover}. In order to synchronise their states, and progress past the intersection, both wagons need to arrive simultaneously. If one wagon arrives before the other, however, then it is almost certain to wander backwards instead of waiting for the other to arrive. If the probability of moving forward is $p$ and backward is $q$, then the system is said to have `computational bias' $b=p-q\ll1$, and the probability that a wagon is at the intersection point is $b/p$. The rate of synchronisation can then be found approximately by the probability that both are there simultaneously, $\sim b^2\lll1$.

  Already the rate of independent computation for each subunit is vanishingly small in the limit of large system sizes, and so synchronisation and communication events are seen to progress even more slowly and appear to `freeze out' as free energy gets sparser and sparser. We show that there is indeed no way to design a better system for synchronisation, evaluating the `time penalty' for a synchronisation event across a broad range of possible designs. We also briefly explain why the same result applies to non-Brownian reversible computers as well, including Quantum computers. We conclude by considering two potential ways to circumvent these issues; the first restricts the quantity of computational subunits permitted to synchronise at any one point, such that they can be granted a greater share of the scarce free energy resource, whilst the second exploits the fact that a Quantum Zeno architecture (described in Part~I) can be operated in a globally serial fashion, in which case synchronisation can be simulated the same way multitasking is simulated on conventional computers.
\end{abstract}

\largertitlepage
\renewcommand{\abstractname}{Technical Abstract}
\begin{abstract}
In Part~I of this series~\cite{earley-parsimony-i}, the limits on the sustained performance of large reversible computers were investigated and found to scale as $\sqrt{AV}$ where $A$ is the convex bounding surface area of the system and $V$ its internal volume, verifying and strengthening a result of \textcite{frank-thesis}, compared to $A$ for an irreversible computer. This analysis neglected to consider interactions between components of the system however, instead focussing on raw computational power. In this part we extend this analysis to consider synchronisation events such as communication between independent reversible processors subject to a limiting supply of free energy. It is found that, whilst asynchronous computation can proceed at a rate $b\lambda$, synchronisation events proceed at the much slower rate $\sim b^2\lambda$; in these rate expressions, $\lambda$ is the gross transition rate for each processor and $b\sim\sqrt{A/V}\ll1$ is the `computational bias' measuring the net fraction of transitions which are successful. Whilst derived for Brownian reversible computers, this result applies to all forms of reversible computer, including Quantum computers. In fact this result is an upper bound, and one must choose the phase space geometry of the synchronisation events carefully to avoid even worse performance. In the limit of large computers, communication will therefore tend to freeze out as $b\to0$; if, however, one is willing to restrict the number of processors permitted to share state at any given time then this rate can be ameliorated and performance on par with asynchronous computation can be recovered.
\end{abstract}

\clearpage

\section{Introduction}


In Part~I of this series~\cite{earley-parsimony-i}, we investigated how the laws of physics constrain the maximum rate of sustained computation within a given region of space. 
Whilst it is known that using reversible computation---computation which conserves information and whose every transition is invertible---allows improved energy efficiency over traditional irreversible computation, we find that the performance of reversible computers scales faster with system size than irreversible computers.
For a region of space of convex bounding surface area $A$ and internal volume $V$, and assuming that the flux of power and entropy per unit surface area is bounded, we found that the rate of an irreversible computer (in units of operations per unit time)
scales proportional to $A$, whilst for a reversible computer it scales proportional to $\sqrt{AV}$.
For a sphere, these are respectively
\begin{align*}
  4\pi R^2&\cdot\frac{\phi}{k_BT\Delta I} & &\text{and}&
  \frac4{\sqrt{3}}\pi R^{5/2}&\cdot\sqrt{\frac{\phi\lambda}{2k_BT}},
\end{align*}
where $R$ is the radius of the sphere, $\phi$ is the power supply per unit surface area, $k_B$ is Boltzmann's constant, $T$ is the surface temperature, $\Delta I$ is the average amount of information erased per irreversible operation, and $\lambda$ is a rate per unit volume of gross reversible computation (whether forwards or backwards). As the reversible rate scales with $R^{5/2}$ compared to the irreversible $R^2$, it will have superior performance above some threshold size (and equal performance below).
This greater bound applied for both classical and quantum computers, although it should be noted that at regions of extreme size---both very small and very large---more restrictive bounds apply; refer to the paper~\cite{earley-parsimony-i} for a more in-depth analysis.

In fact, this performance measure of operations per unit time is not the only one of interest. As we are considering computers of arbitrary size, interaction between different subvolumes will often be of considerable import. In this Part and Part~III, we evaluate performance within this context. Part~II concerns the statistical dynamics of communication between two or more distinct computational units, whilst Part~III concerns the statistical dynamics of computational units interacting with resources or chemical species, of which there may be a variable number of identical units or particles. For convenience, we shall call particles which are uniquely distinguishable or addressable \emph{mona} (sg.\ \emph{monon}), after the Greek for unique, \emph{\gr μοναδικός}. Similarly, we will call indistinct particles belonging to a species \emph{klona} (sg.\ \emph{klonon}). Klona may be likened to traditional chemical species, such as glucose and ATP. Mona are less commonly considered; an example would be a 
\SI{150}{\milli\liter} solution containing a \SI{1}{\milli\molar} concentration of random nucleic acids of length 80 nucleotides\footnotemark. 
We can therefore say that Part~II studies mona-mona interactions, and Part~III mona-klona interactions. We do not cover klona-klona interactions as these are well understood within the domains of chemistry and statistical physics.
\footnotetext{A nucleic acid, such as the DNA molecules that compromise our genome, is a polymer formed from an arbitrary sequence of four nucleotide monomers; random nucleic acids are easily obtained from DNA synthesis companies. As such, there are $4^{80}\approx\num{1.5e48}$ possible distinct 80-mers (nucleic acid of length 80). The birthday paradox says that we need $\num{1.4e24}$ of these random molecules before it is likely that there are duplicates present. Our \SI{150}{\milli\liter}, \SI{10}{\milli\molar} solution contains just shy of $\num{e20}$ such molecules, and it can be calculated that the probability that there are any duplicates in the solution is less than 1 in 350 million. Therefore, it is almost certain that every nucleic acid in said solution is a monon.}


The results in this paper will apply to generic models of computation, but for the sake of illustration it is useful to introduce a canonical computational system.
We shall use Brownian computers, following the formulation given in Part~I;
in particular, we will imagine a bag of water in which are dissolved various chemical species. Included among these species will be our mona, molecular computers. Examples of molecular computers can be found in the field of DNA computing~\cite{rothemund-tm,winfree-tam,qian-seesaw,qian-sqrt,crn-rm,dsd,cardelli-dsd,qian-nn,winfree-dna-crn-univ,pen}, wherein the ability for nucleic acids to bind to other nucleic acids in a sequence-specific manner\footnotemark\ is exploited in various different ways to program cascades of reactions encoding arbitrary computation. To take full advantage of such molecular computers we need to let them communicate and interact.
\footnotetext{Nucleic acids are polymers of arbitrary sequences of four monomers. In the case of DNA, these monomers are adenine, guanine, cytosine and thymine or \texttt{A}, \texttt{G}, \texttt{C} and \texttt{T}. The nucleotide monomers of nucleic acids are special in that they are subject to base-pairing rules: \texttt{A} preferentially binds to \texttt{T} and vice-versa, and \texttt{G} preferentially binds to \texttt{C} and vice-versa. The consequence is that a DNA sequence \texttt{5'-CTTGCATCGC-3'} will generally only bind stably to its (reverse) complementary sequence \texttt{5'-GCGATGCAAG-3'}.}

Recalling that we are considering interactions between \emph{mona} in this paper, each interaction event will be between specific mona. If these mona diffuse freely throughout the volume, then communicating with a specific monon will be impracticable except in the case of very small (effective) volumes. On average, the time between collisions of $n$ such mona in a system of volume $V$ will scale as $\sim\bigOO{V^{n-1}}$. Moreover, we shall see that a significant problem in reversible communication is that even after mona meet they can go backwards in phase space at which point we must wait for them to collide again; clearly in the freely diffusive case this is untenable. To improve upon this scaling, we are forced to introduce a lattice-like structure into the system. This lattice should be endowed with a coordinate system and the coordinates should be logically chosen so that it is `easy' to deterministically compute a unique path between any pair of reachable coordinates. Interacting mona should bind to this lattice, and couple their movement along it to their computational bias (\Cref{dfn:bias}, below). In so doing, translocation along the lattice becomes just another part of the monon's computational state---the information content of the monon that may be transformed in the course of computation.

\begin{dfn}[Computational Bias]\label{dfn:bias}
  Recall from Part~I~\cite{earley-parsimony-i} that in a Brownian computer, transitions are mediated by systems of klona representing `positive' and `negative' bias. Physically, such klona act as a source of free energy and a real world example is given by $\ce{ATP}$ and $\ce{ADP + P_{\textrm i}}$ in biochemical systems. In the simplest case, there are $\oplus$ and $\ominus$ klona such that $\forall n.~\ce{$\ket{n}$ + $\oplus$ <=> $\ket{n+1}$ + $\ominus$}$ where $\ket{n}$ is the $n^{\text{th}}$ state of some monon, and this is positively biased when the concentration of $\oplus$ exceeds that of $\ominus$. We define the bias in this case to be $b=([\oplus]-[\ominus])/([\oplus]+[\ominus])$.
\end{dfn}

\begin{figure}
  \centering
  \begin{subfigure}{.58\linewidth}
    \includegraphics[width=.9\linewidth]{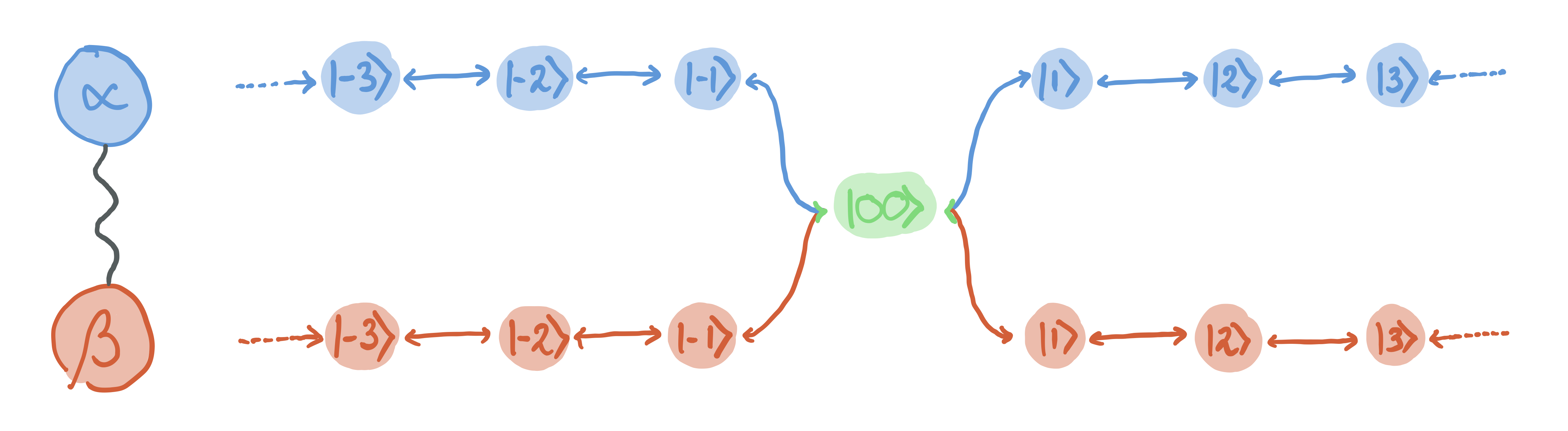}
    \caption{}\label{fig:synch-juxt}
  \end{subfigure}

  \begin{subfigure}{.207\linewidth} 
    \includegraphics[width=.97\linewidth]{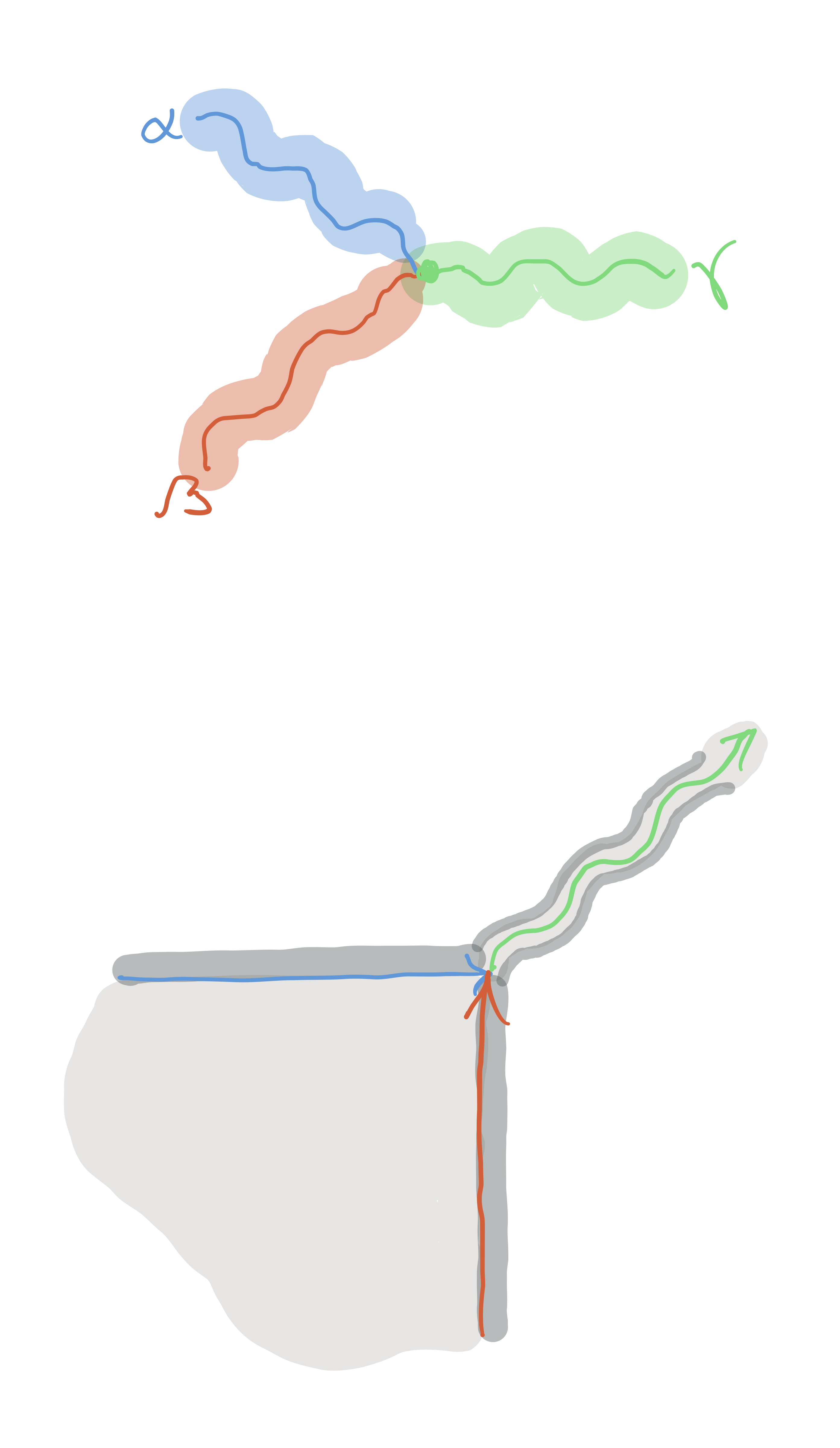}
    \caption{}\label{fig:synch-simple-1}
  \end{subfigure}%
  \begin{subfigure}{.228\linewidth} 
    \includegraphics[width=.97\linewidth]{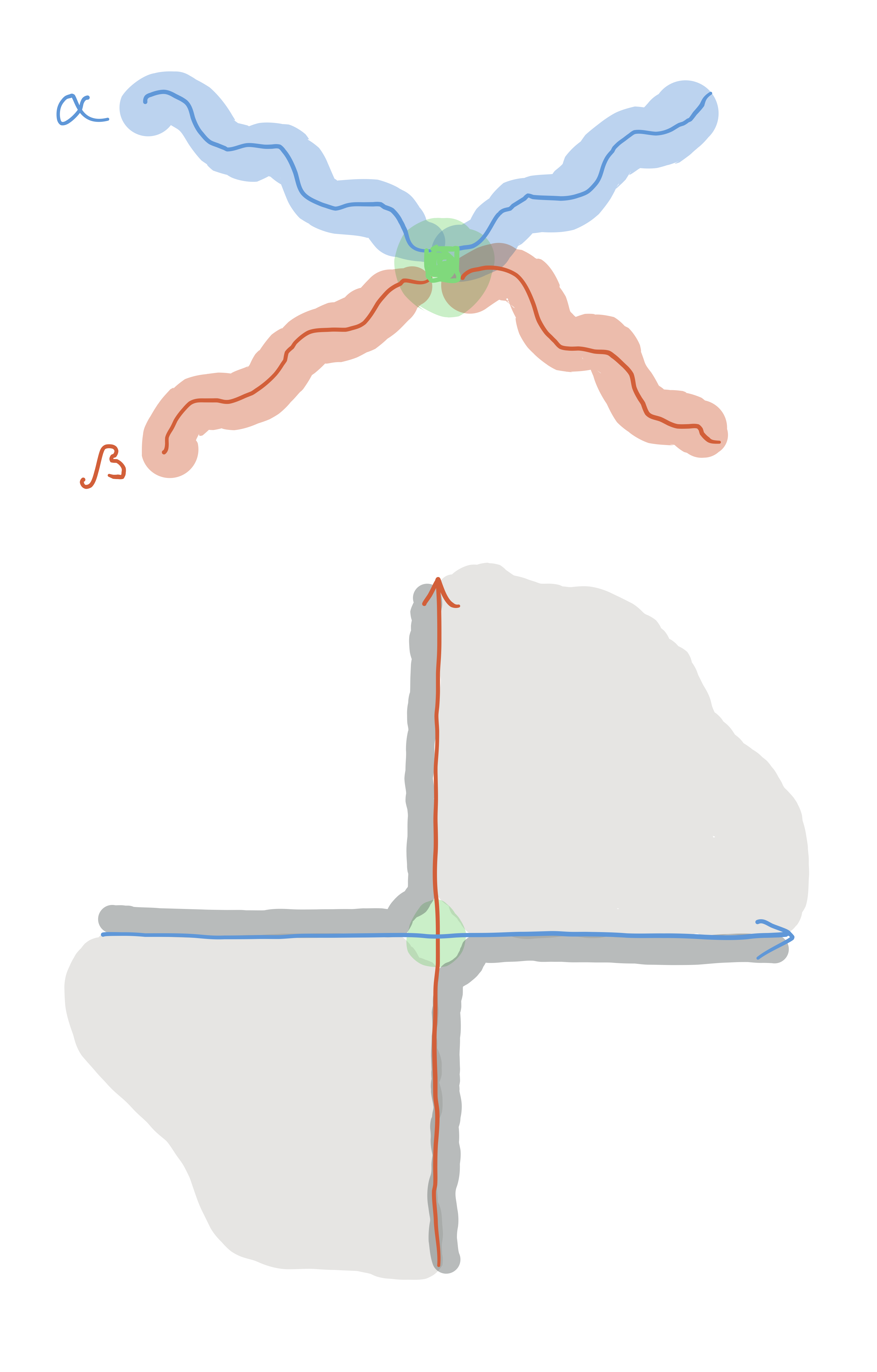}
    \caption{}\label{fig:synch-simple-2}
  \end{subfigure}%
  \begin{subfigure}{.284\linewidth} 
    \includegraphics[width=.97\linewidth]{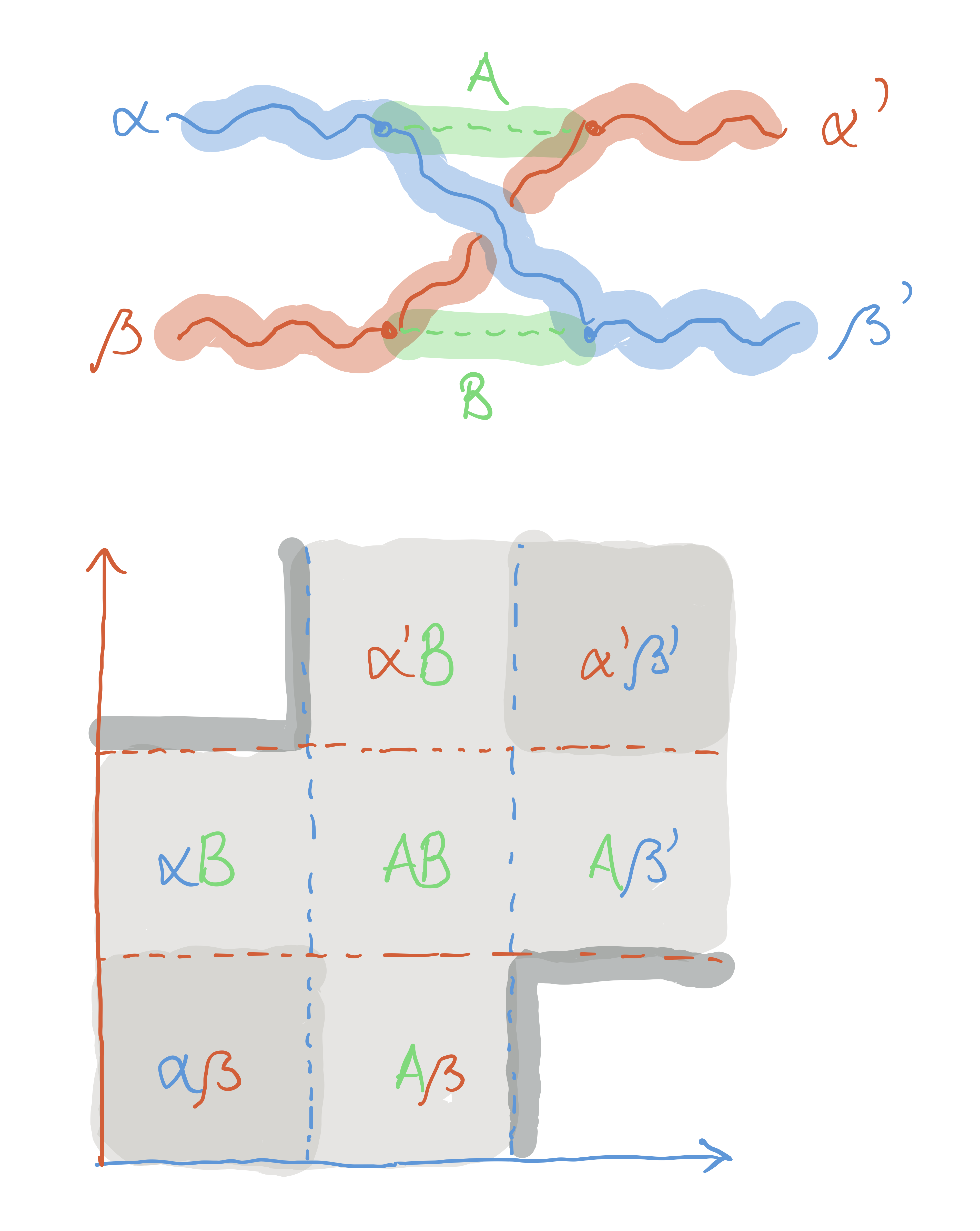}
    \caption{}\label{fig:synch-wide-1}
  \end{subfigure}%
  \begin{subfigure}{.260\linewidth} 
    \includegraphics[width=.97\linewidth]{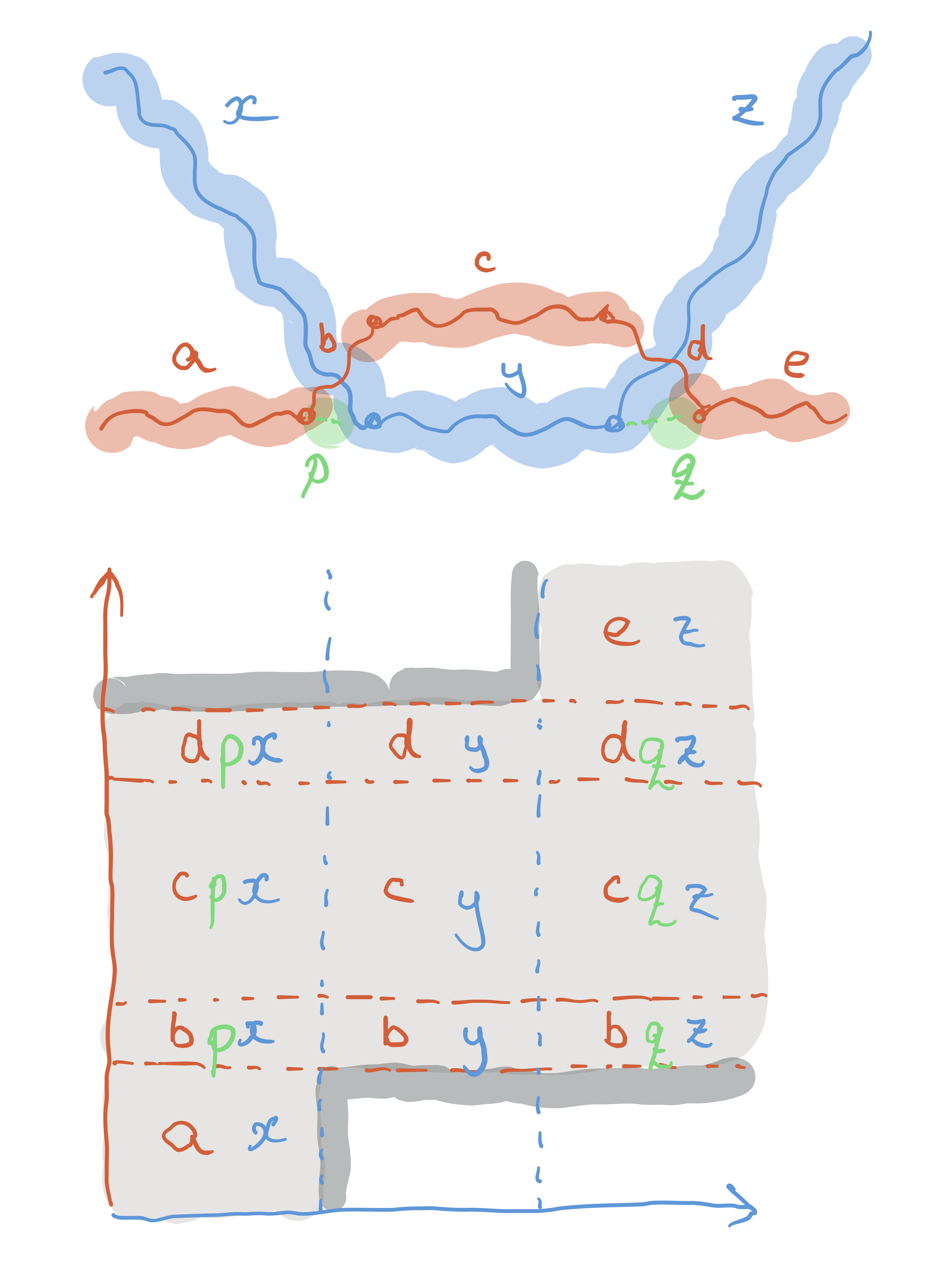}
    \caption{}\label{fig:synch-wide-2}
  \end{subfigure}

  \begin{subfigure}{.316\linewidth}
    \includegraphics[width=.97\linewidth]{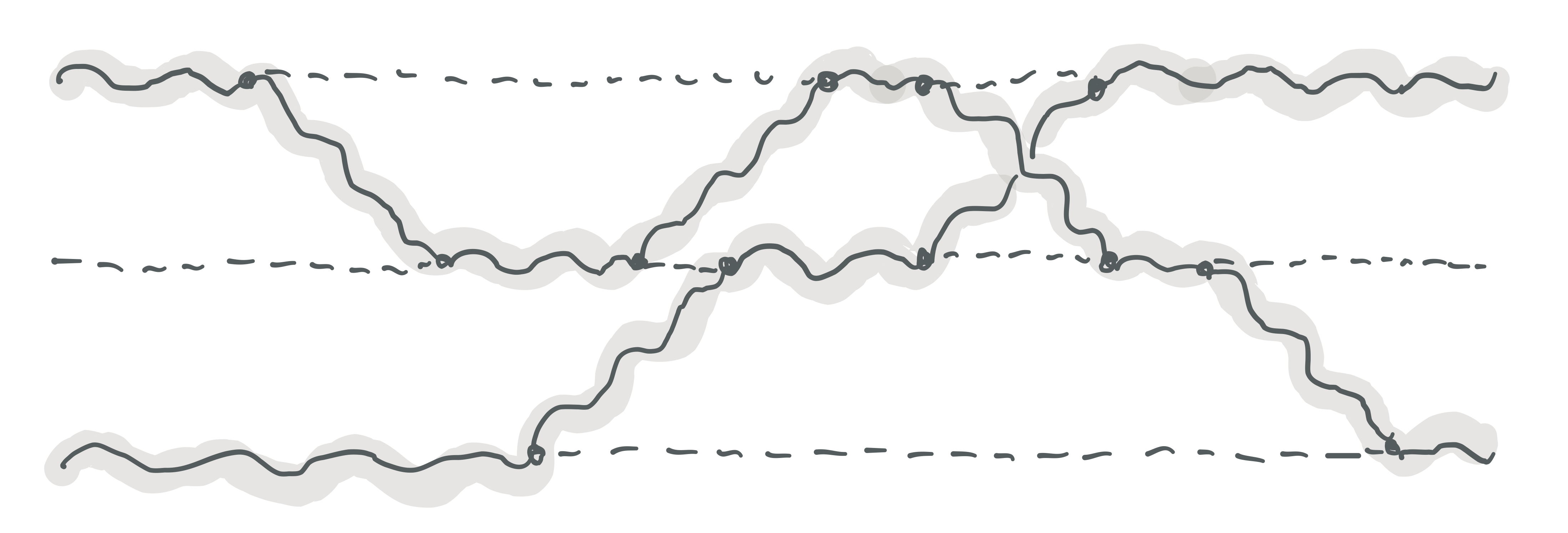}
    \caption{}\label{fig:synch-multi-1}
  \end{subfigure}%
  \begin{subfigure}{.240\linewidth}
    \includegraphics[width=.97\linewidth]{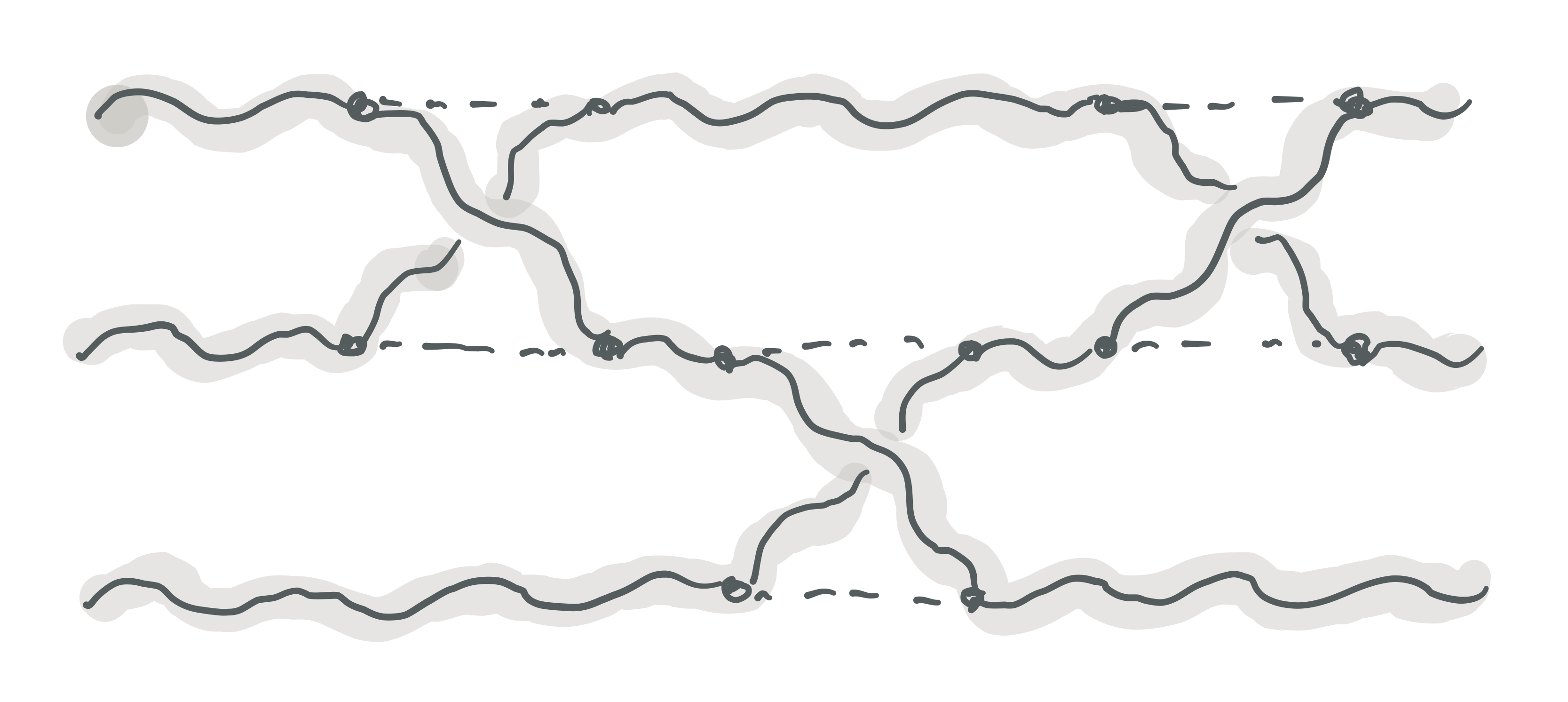}
    \caption{}\label{fig:synch-multi-2}
  \end{subfigure}

  \caption{A range of examples of \emph{constrictive} synchronisation problems, represented in different ways. Figure (a) shows a state diagram for two bound mona $\alpha$ and $\beta$; each monon is free to walk back and forth along their state space ($\mathbb{Z}$) independently except for at their \emph{synchronisation point}, $\ket{00}$, through which they must pass together. Figures (b--e) show a range of increasingly complex examples from two perspectives; the top diagrams are abstract state diagrams reminiscent of Feynman diagrams, whilst the bottom diagrams show the joint phase space of the mona. Dashed lines in state space, or `artefacts', represent static/non-evolving mona; for example, in (d) monon $\alpha$ leaves an artefact $A$---which may contain some informational payload or data packet---that is then absorbed by monon $\beta$, and vice-versa for $B$. More specifically, an artefact is a packet of data left on the lattice by a computational monon after it departs, ready to be absorbed by another computational monon.\\
  \hspace*{1em}Figure (b) is an asymmetric fusion/fission synchronisation, depending on the direction of time. Figure (c) is symmetric, and equivalent to (a). Figure (d) shows an approach to `widen the constriction point' of (c) using artefacts, thus making synchronisation easier. Figure (e) shows how two synchronisations close in phase space lead to an apparent `tunnel'. Finally, figures (f--g) show some more complicated examples suggestive of what a real reversible communicating system might look like.}\label{fig:synch-ex-1}
\end{figure}

\begin{figure}
  \centering
  \begin{subfigure}{.24\linewidth}
    \includegraphics[width=.97\linewidth]{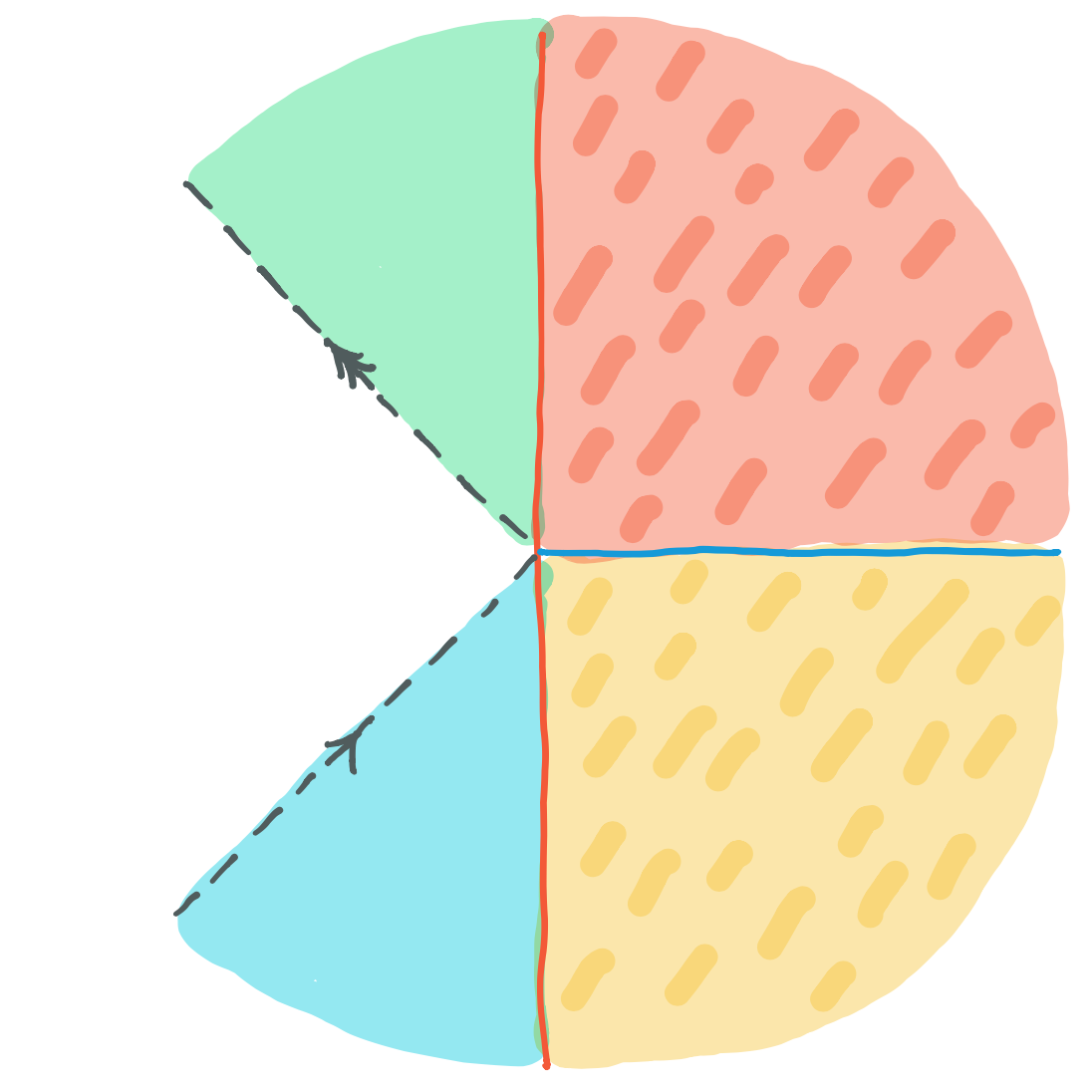}
    \caption{}\label{fig:synch-rec-1}
  \end{subfigure}~~%
  \begin{subfigure}{.24\linewidth}
    \includegraphics[width=.97\linewidth]{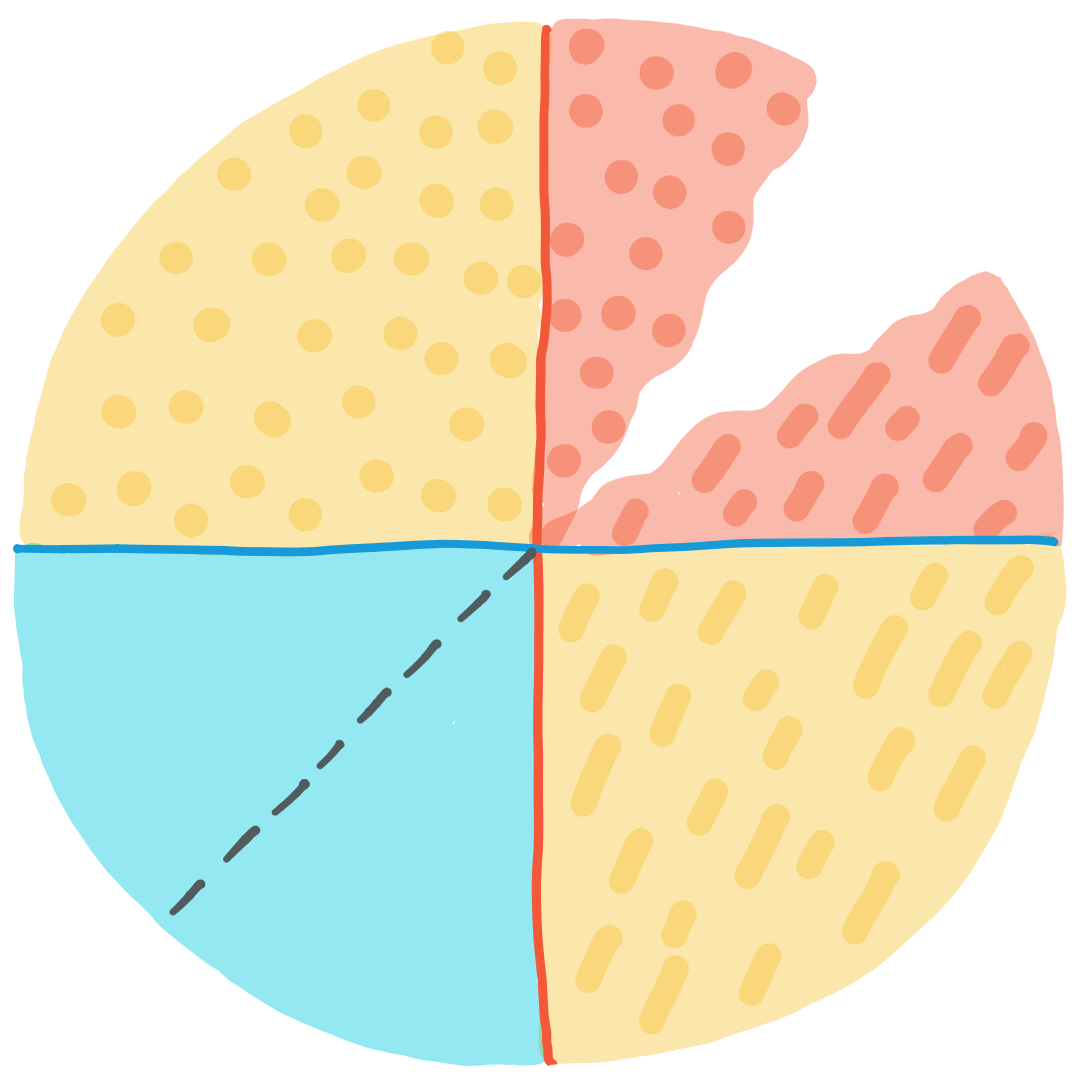}
    \caption{}\label{fig:synch-rec-domain}
  \end{subfigure}~~%
  \begin{subfigure}{.24\linewidth}
    \includegraphics[width=.97\linewidth]{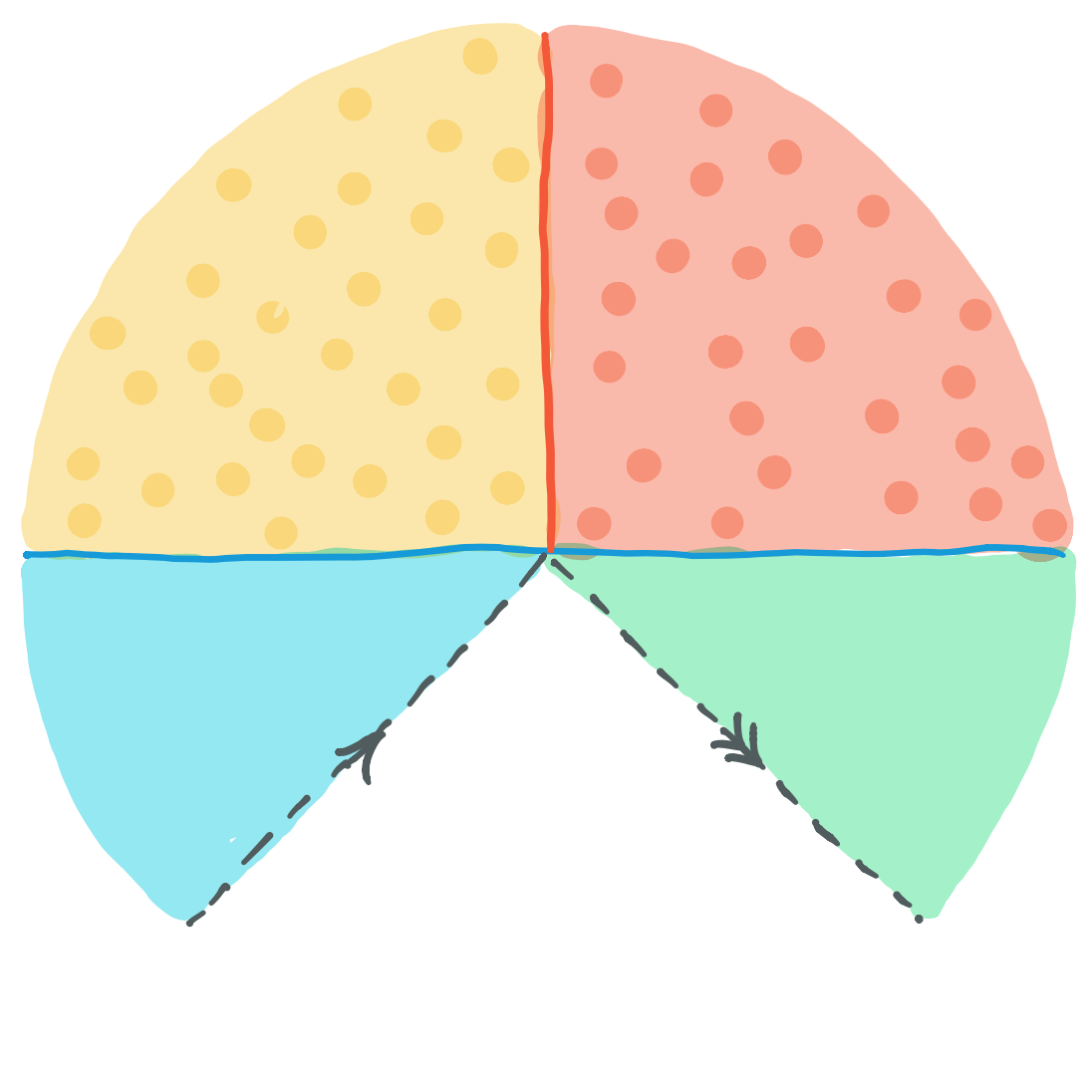}
    \caption{}\label{fig:synch-rec-2}
  \end{subfigure}
  \\[0.4cm]
  \begin{subfigure}{.3\linewidth}
    \includegraphics[width=.97\linewidth]{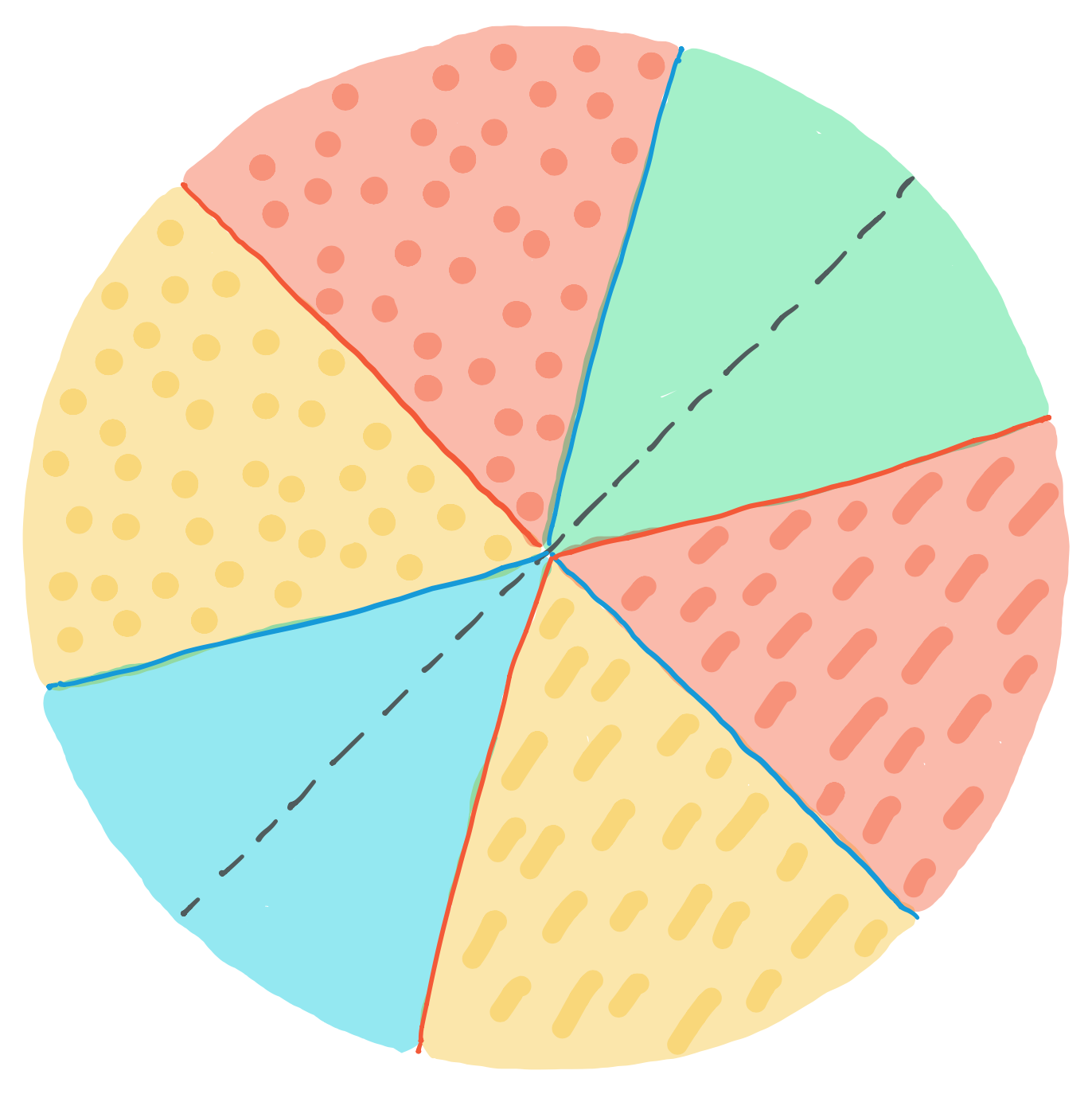}
    \caption{}\label{fig:synch-rec-total}
  \end{subfigure}

  \caption{An example of a recessive synchronisation problem using a phase space depiction. This synchronisation concerns two mona, say $\alpha$ and $\beta$ (blue and red axes respectively), starting in the blue quadrant and ending in the green quadrant. In contrast to constrictive synchronisations as illustrated in \Cref{fig:synch-ex-1}, a recessive synchronisation avoids restricting which transitions are permissible anywhere in phase space. As such, locally all four transitions are allowed (two for each monon, backward and forward). In order to achieve this, we must deal with the case wherein one monon, say $\alpha$, reaches the synchronisation point before the other, $\beta$. In order to not have its forward transition blocked, we introduce `dummy' evolution whilst it waits for $\beta$ to arrive. This dummy evolution is essentially a counter over $\mathbb{N}$ of how long $\alpha$ has been waiting for $\beta$, and corresponds to the yellow quadrant (there are two yellow quadrants, one for $\alpha$ waiting and one for $\beta$). When $\beta$ arrives, the two finally synchronise. Before $\alpha$ can continue onward though, this counter must now be erased; we do so reversibly by essentially reversing the blue-yellow process to obtain a red-green process, which counts down. When the counter reaches 0, $\alpha$ continues. This is a reversible erasure because the final count of $\alpha$ before synchronisation is not important, and so if the bias was reversed---to go from green to red to yellow to blue---then we may reach a different counter value whilst we wait to rendezvous with $\beta$. Counting all the regions, we have a total of six quadrants. Diagrams (a--c) show phase subspaces without distortion, and can be compared to Riemann sheets. Diagram (a) shows the subspace corresponding to $\alpha$ reaching the synchronisation point first, and (c) corresponds to $\beta$ reaching it first. The blue and yellow sectors of Diagram (b) is the pre-synchronisation region for this problem. Gluing them together along the dashed lines reveals the complete Riemann surface, Diagram (d), but must be distorted in order to project it onto the two dimensional Euclidean surface of this paper.}\label{fig:synch-ex-2}
\end{figure}

In order for two or more mona to communicate, or more generally to synchronise their joint state, we must arrange for their states to coincide at some point. Moreover, no monon will be able to proceed past this synchronisation point without the other(s). In the irreversible limit of computing, this problem is inconsequential as each computational entity can simply wait for the others. In reality, `waiting' is not a physically reversible process: If the dynamics were reversed it would not be possible for the entity to `know' how long to wait before going backwards. In an irreversible system, waiting can be simulated by raising the entropy of the environment; for example, the system could continually `forget' how long it has been since it arrived by resetting some internal state using the Landauer-Szilard principle~\cite{szilard-engine,landauer-limit}, or the system could transfer its generalised computational `momentum' to some other computational entity. In the reversible computers we are considering, there is a dearth of free energy\footnotemark\ as we are distributing the free energy supplied at the computer's surface throughout its entire volume, and so these approaches are not generally possible. 
\footnotetext{Free energy is the ability of the system to drive an entropically unfavourable reaction, by increasing the entropy of the environment at least as much as the reaction reduces the local entropy.}

Instead, the simplest approach is to let the mona `bounce off' the synchronisation point until the other arrives, relying on the (weak) computational bias to drive them towards this. That is, upon reaching the synchronisation point the monon cannot proceed and has a high chance of walking backwards, but the weak bias will keep it within a loose vicinity of the synchronisation point. Some initial estimates (\Cref{sec:constrict-initial}) will show that, in addition to the time for each monon to reach the synchronisation point independently, there is a `penalty' term. For computational bias $b$, each net computational step for lone mona takes time $(b\lambda)^{-1}$ where $\lambda$ is the rate at which mona collide with bias klona. If each participating monon starts at a phase distance $n_i$ from the synchronisation point, then the expected time for all to have briefly reached the synchronisation point is $\tau_0=(b\lambda)^{-1}\max_i n_i$. The penalty depends on the joint probability distribution of the mona, and how often they are all found at the synchronisation point. Analysing this in detail turns out to be surprisingly difficult, but we will succeed in showing a range of approximate, empirical and exact results, confirming the initial estimate of a penalty $\Delta\tau\gtrsim (b^d\lambda)^{-1}$ for $d$ mona. As $b\ll1$ generally, this will be very large compared to the timescales of independent computation by individual mona, and thus potentially rendering reversible communication impracticable.

In this paper, we shall study this problem in detail for general shapes of synchronisation phase spaces and illustrate with examples of the forms of phase space that would be commonly found in reversibly communicating mona. We divide approaches to synchronisation in two: \Cref{sec:constrict} will concern the `constrictive' case in which there is a constriction in phase space, as illustrated in \Cref{fig:synch-ex-1}. \Cref{sec:recessive} will concern the `recessive' case in which the phase space is not restricted, and therefore other methods must be used to synchronise the joint state of the mona as illustrated in \Cref{fig:synch-ex-2}. In fact, there are two more related cases, dilatory and processive, which are in a sense complementary to the constrictive and recessive cases respectively. In the dilatory case, the phase space widens towards the synchronisation point rather than shrinking, whilst in the processive case the boundary of the post-synchronisation subspace (green quadrant in \Cref{fig:synch-ex-2}) is concave rather than convex. 
In both of these, the penalty will be negligibly small, zero, or even negative as the phase coordinate is pushed entropically towards larger regions of phase space. Nevertheless, they are not desirable because such asymmetry implies an increase in entropy and hence irreversibility. We can therefore make a stronger and more accurate statement: rather than forbidding dilatory cases, we require synchronisation phase spaces to be symmetric in the synchronisation point or subspace.

\section{Methodology}\label{sec:methodology}

\para{Isolated Mona}

\begin{figure}
  \centering
  \begin{subfigure}{.8\textwidth}
    \includegraphics[width=\linewidth]{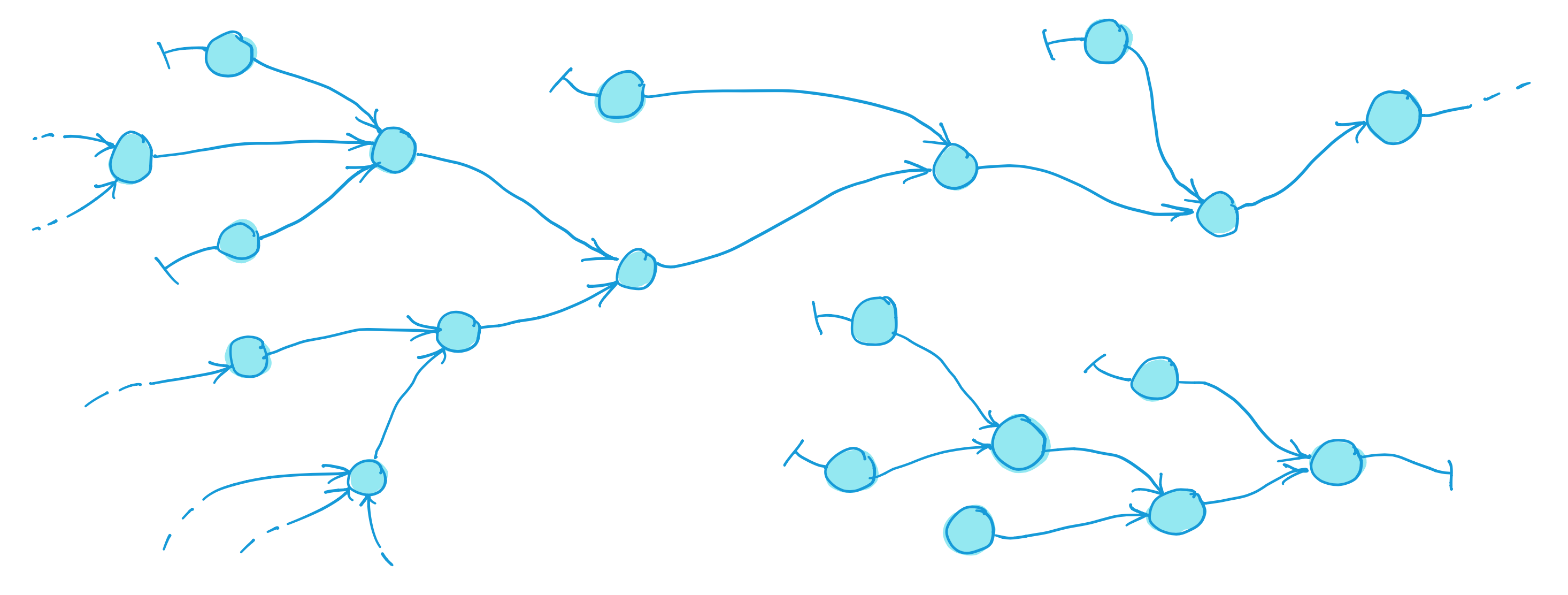}
    \caption{Examples of irreversible state spaces}
  \end{subfigure}
  \begin{subfigure}{.7\textwidth}
    \vspace{10pt}
    \includegraphics[width=\linewidth]{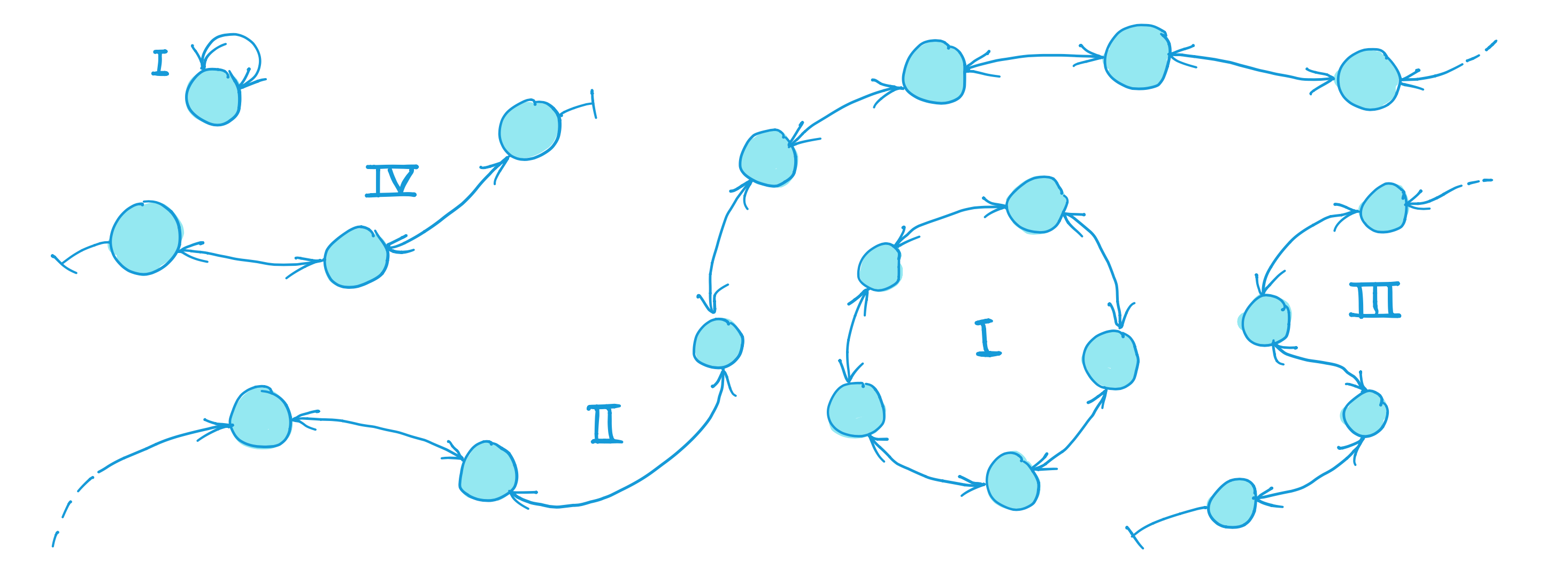}
    \caption{Examples of reversible state spaces}
  \end{subfigure}
  \caption{These figures show some examples of the state spaces of (a)~irreversible and (b)~reversible computers. The blue nodes represent distinct computational states, and arcs represent computational transitions or operations. Terminal arcs, represented by $\vdash$, indicate that computation either begins or ends here; that is, no predecessor or successor states (respectively) exist. Recall that the dynamics of the system are free to wander forwards and backwards in state space, including away from a terminal state. The distinction between irreversible and reversible computers is that an irreversible state may (and often \emph{does}) have more than one predecessor state. Where an irreversible computer completely forgets or discards a variable---such as of type \texttt{uint32}---it may have a very large number of predecessor states---for \texttt{uint32}, $2^{32}\approx\num{4e9}$. For abstract models of computation such as a Register Machine (see, e.g., \textcite{register-machine}), there can even be a countable infinity of predecessors in the case of the `clear-to-zero' operation. In contrast, a reversible state may have at most one predecessor and one successor state. We can classify reversible computers into four classes; (I) finite cyclic programs with no termini, (II) bi-infinite programs with no termini, (III) infinite programs with one terminus (comparable to non-halting programs in computability theory), (IV) bi-terminating programs with two termini (i.e.\ finite linear chains). Bear in mind that the form of this state space is separate from that of the program logic: a reversible program may well contain loops and conditional branching, but for a given program and input or output, evolution is deterministic both forwards and backwards.}
  \label{fig:state-space}
\end{figure}

To formalise the problem statement, we begin by considering computation by isolated mona in more detail. A monon's evolution is best understood by considering the state space of the computation it is simulating as depicted in \Cref{fig:state-space}. Ignoring any possible intermediate states, there will be an isomorphism between the states of the monon and the states of its underlying computation. As was made apparent in our previous paper~\cite{earley-parsimony-i}, we will want to drive computation forward by coupling the monon's state transitions to some non-equilibrium system of klona, which we term a bias source. Each transition may be coupled to a different bias source, and the strengths of the bias sources may vary over time. The definition of bias sources and their strengths is recalled in \Cref{dfn:bias}. Where possible, we will remain general in our analysis; however, our previous paper showed that optimal performance is achieved for bias constant in time and uniform in space, and our illustrative examples will reflect this fact.

The dynamics of this system can therefore be seen to be a walk in a 1-dimensional phase space where at any point, there are at most two transitions available: one to the successor state, and one to the predecessor state, should they exist. Depending on whether we use a model with discrete or continuous time\footnotemark, these transitions will be labelled with probabilities or rates that depend on the strength of the bias source(s). We note that we expect the strength of the bias sources to be very small, such that the timescale of making one step of net progress is very large compared to that of a single transition. As a result, the monon will experience a time-averaged bias, and any correlations in the system will be expected to have been exponentially suppressed by diffusive processes. Therefore, we can model the monon's dynamics as a Markov process or chain; namely, the dynamics of the system depend only on its current state, rather than its history.

As mentioned in the introduction, we shall incorporate spatial position as just another part of our computational state, such that translocation along a lattice (or indeed any other environmental interaction) can be treated as just another computational transition. This abstraction allows us to greatly simplify our analysis, as we need only track position in phase space; the exact meaning of different phase coordinates is irrelevant, and so our results will remain general in a large class of synchronisation processes.
\footnotetext{Continuous time is the more apt model, however for generating function approaches discrete time must be employed. Fortunately, when the bias rates are uniform, it is easy to convert between the two by simply dividing the times by the gross bias rate, $\lambda=\lambda_\oplus+\lambda_\ominus$.}

\para{Interacting Mona}

For a system of multiple particles, the state of the system is given by a coordinate in their joint phase space. In a classical physical system of $N$ particles, this manifests as a $6N$-dimensional phase space as each particle has six degrees of freedom: three for the particle's position, and three for its momentum vector. Similarly, our system of $N$ mona has as states coordinates in an $N$-dimensional phase space. When these mona do not interact, the phase space naturally decomposes into a tensor product of each monon's individual phase space. If, however, $n$ of these mona happen to interact with one another, then the subspace formed from their degrees of freedom is irreducible: it cannot be further decomposed into a tensor product. The reason is that a synchronisation manifests as a constraint on the joint state.

To make this more concrete, suppose three mona are to share some information. Let their individual states be indexed by $\mathbb Z$, such that their joint phase space is given by a subset of $\mathbb Z^3$. Suppose the interaction occurs when each monon is in state 0, i.e.\ the coordinate at which they interact is $(0,0,0)$. As the future state of each monon depends on this interaction/communication/synchronisation event, we observe that some coordinates are causally impossible. For example, $(3,-4,5)$ would imply that the first and third mona have received information from the second, despite the second having never given them this information. In fact, we find that the only valid states are given by $\{(-a,-b,-c):a,b,c\in\mathbb N\}\cup\{(a,b,c):a,b,c\in\mathbb N\}\equiv(-\mathbb N)^3\cup\mathbb N^3$. This corresponds to the negative and positive octants of $\mathbb Z^3$, and the system must pass through the origin. The equivalent case of two mona is depicted in \Cref{fig:synch-juxt,fig:synch-simple-2}.

This is not the only way for mona to interact as illustrated by other examples in \Cref{fig:synch-ex-1}. The common factor is that a synchronisation event between mona with individual phase spaces $\mathcal R_i$ corresponds to their joint state $\mathcal R$ being a strict subset of $\mathcal R' = \medotimes_i\mathcal R_i$. Our aim is to analyse the mean time to get from some region of phase space to another, and for this we shall need to specify some regions: $\mathcal I$ will be the initial region of phase space (or more generally, a distribution over phase space), $\mathcal P$ will be the region `before' synchronisation, $\mathcal P'$ the region `after', and $\mathcal S$ the interfacial region corresponding to the process of synchronisation. These phase regions must satisfy the relations $\mathcal I \subset \mathcal P$, $\mathcal P \cap \mathcal P' = \mathcal S$, and $\mathcal P \cup \mathcal P'=\mathcal R$. It should be noted that the synchronisation region $\mathcal S$ is arbitrary, although in practice there is usually a natural choice.

The question naturally arises of what happens at the boundaries of this phase space. Quite simply, transitions which would take the mona outside of the valid phase region are blocked; that is, there is no reaction with reactants $\ket{0}$ and $\oplus$ if the other mona are not all receptive to synchronisation. A more abstract but more general answer is that the boundaries of $\mathcal R$ are \emph{reflective}: the probability flux/current normal to them vanishes.

We now introduce a canonical slicing of $\mathcal P$ into hypersurfaces $\{\mathcal P_s:s\in\mathbb N\}$. The hypersurface $\mathcal P_s$ is given by the locus of nodes whose shortest path (i.e.\ the minimum number of transitions) to $\mathcal S$ is $s$. By definition, $\mathcal P_0\equiv\mathcal S$, and it can be shown that to get from a node in $\mathcal P_{s+k}$ to $\mathcal P_s$, one must pass through the hypersurfaces $\{\mathcal P_{s+k-1},\mathcal P_{s+k-2},\ldots,\mathcal P_{s+1}\}$ in turn. We will usually take $\mathcal I\subseteq\mathcal P_s$ for some `distance' $s$.

These definitions allow us to categorise the forms of synchronisation events. If $|\mathcal P_{s+1}| \ge |\mathcal P_s|$ for all $s$ within an appropriate vicinity of $\mathcal S$ where $|\mathcal P_s|$ is the number of nodes in said hypersurface, then we call this a constrictive synchronisation. If instead $|\mathcal P_{s+1}| \le |\mathcal P_s|$ we call it dilatory. These categories intersect in the special case of $|\mathcal P_s|$ constant; this can be likened to a tube, and has zero penalty. For uniformly positive bias, a dilatory synchronisation problem does not present an impediment to synchronisation and in fact may even aid the process, as the reflective boundary serves to `push' probability density towards $\mathcal S$. In practice, dilatory synchronisations only occur in contrived scenarios, if at all, and synchronisations can be assumed constrictive. If a synchronisation geometry does not fit into either of these categories, then it may need to be partitioned into constrictive and dilatory regions, each of which can be analysed separately. Alternatively, if the geometry is `on average' constrictive then one should be able to cautiously apply the relevant results.

\para{Locally Unconstrained Synchronisation}

As alluded to in the introduction and \Cref{fig:synch-ex-2}, a synchronisation problem can be modified to `fill in' the missing regions of phase space. This is achieved by introducing a `dummy' counter for each monon. Whenever a monon reaches the synchronisation subspace (i.e.\ it is receptive to synchronisation) it begins counting the time it has waited. This is not quite accurate, as the counter is just another part of the monon's computational state and so it can go down as well as up; nevertheless, on average it counts up. The consequence is that every monon can continue to evolve unimpeded. Once the last monon becomes receptive to synchronisation, the mona finally synchronise. At this point, the latent mona begin to decrement their counters. When a monon reaches zero, it may resume its original computation. Though a subtle point, this is indeed a reversible computation despite appearing to erase information, and is made clearer in \Cref{fig:synch-ex-2}. 

In return for the elimination of phase space boundaries, the phase space becomes effectively larger, and control over the state of the mona is more limited. In particular, with a weak bias the mona are free to wander into these dummy regions. We will analyse this case in \Cref{sec:recessive} in order to compare its performance to the constrictive case. As with the constrictive/dilatory cases, we can classify these problems into recessive and processive cases. Recessive cases correspond to $\mathcal P'$ being convex, and processive to it being concave. Again, it is not obvious how a processive case could be constructed, if indeed it can be. Moreover, it is unclear how to analyse a case that does not fall into these categories.

\para{Problem Statement}

With the synchronisation phase space well defined, we are now in a position to formulate a question. Our goal is to determine the mean time to get from $\mathcal I$ to $\mathcal S$, and to compare this to the unsynchronised system $\mathcal R'$. For homogeneous constant bias $b$, it is straightforward to show that unconstrained evolution from $\mathcal P_s$ to $\mathcal P_0=\mathcal S$ takes mean time $\tau=s/b\lambda$. That is, the computational 'speed' is $1/b\lambda$. More generally, we can define this quantity by $\tau = \inf\{t:\forall u>0.\ev{x(t+u)}\in\mathcal P'\}$. That is, it is the least time for which the mean position in phase space never leaves the post-synchronisation domain. For many systems, this can be identified with the first time at which the mean position in phase space enters the post-synchronisation domain, $\tau = \inf\{t:\ev{x(t)}\in\mathcal P'\}$.

To compute this latter quantity $\tau$ in general cases, we consider the evolution of the phase space probability distribution, $W(t)=W(\vec x;t)$. This quantity $\tau$ is well known in the literature, and is termed the Mean First-Passage Time, or MFPT. We will generally follow the approach of \textcite{risken}, in particular Chapter~8. The first observation we can make is that we need not track particles that cross the boundary between the two domains; that is, we can place an absorbing boundary at $\mathcal S$. The total extant probability, $G(t)=\int_{\mathcal P\setminus\mathcal S}\dd{\vec x}W(t)$, can then be interpreted as a survival probability, $\pr(\tau>t)$. Introducing a probability density $\varrho(t)$ for the MFPT, such that $\tau=\int_0^\infty\dd{t}t\varrho(t)$, we can see that $G(t)=\pr(\tau>t)=\int_t^\infty\dd{t'}\varrho(t')$ and hence $\tau=-\int_0^\infty\dd{t}t\dv{G}{t}=\int_0^\infty\dd{t}G(t)$.

We proceed by reducing the dimension of the problem. Surprisingly, it transpires that we can in fact eliminate the time variable. We recall the standard approach from \textcite{risken}. The $n^{\text{th}}$ moment of $\tau$ is found to be
\begin{align*}
  \ev{\tau^n} &= -\int_0^\infty \dd{t}t^n\pdv{t}\int_{\mathcal P}\dd{\vec x}W(\vec x;t)=\int_{\mathcal P}\dd{\vec x}\underbrace{\left(-\int_0^\infty\dd{t}t^n\pdv{t}W(\vec x;t)\right)}_{w_n(\vec x)}.
\end{align*}
Let $\mathcal L=\mathcal L(\vec x)$ be an operator describing the time evolution of the phase distribution, i.e.\ $\dot W=\mathcal LW$. Here we require that $\mathcal L$ not depend on time, and so we will not consider time-variable biases from here on. Integrating by parts and then applying $\mathcal L$, we can relate the $w_n$ to each other:
\begin{align*}
  w_{n+1}(\vec x) &= n\int_0^\infty\dd{t}t^nW(\vec x;t), \\
  \mathcal Lw_{n+1}(\vec x) &= n\int_0^\infty\dd{t}t^n\mathcal LW(\vec x;t) \\
    &= n\int_0^\infty\dd{t}t^n\pdv{t}W(\vec x;t) \\
    &= -nw_{n-1}(\vec x).
\end{align*}
Evaluating the base case, $w_0(\vec x)=W(\vec x;0)$, we thus have an inductive definition of the $w_n$, from which we can obtain each moment of $\tau$. As these turn out to be non-trivial to evaluate, we will focus on the first moment---the mean---in this paper, i.e.\ $\tau=\int_{\mathcal P}\dd{\vec x}w_1$ with $\mathcal Lw_1=-W(0)$ where $W(0)$ is the initial distribution on $\mathcal I$. That is, we merely need to compute the steady state distribution of the system $\mathcal P$ with absorbing boundary at $\mathcal S$, subject to forcing $W(0)$. Note that, whilst $W(0)$ is a normalised probability distribution with total density 1, $w$ is not. In fact, its total density is precisely the MFPT. 

A useful observation is that phase density is conserved by the operator $\mathcal L$. This means we can write $\mathcal L=-\vec\Delta\cdot\vec S$ for discrete phase space, where $\vec\Delta$ is the discrete vector derivative (or finite difference) and $\vec S$ is the phase current. For continuous phase space, we have instead $\mathcal L=-\vec\nabla\cdot\vec S$. Current will be very important to our analysis, and is informed by the sources and sinks due to the $W(0)$ forcing and absorbing boundary respectively.

\subsection{Techniques for Discrete Phase Space}

\para{Recurrence Relations}

There are a number of ways to represent the operator $\mathcal L$ for a discrete system. The canonical approach to describe a Markov chain is by introducing a distribution vector with indices corresponding to loci in phase space. For example, the phase space corresponding to \Cref{fig:synch-juxt} would be represented by a vector in the infinite dimensional vector space indexed by the set $\{(-i,-j):i,j\in\mathbb N\}\cup\{(i,j):i,j\in\mathbb N\}$. The operator is then an infinite matrix for this space. Whilst a convenient approach due to the wealth of techniques built around this representation, in addition to the ability to solve for the steady state as an eigenvector problem, it is too unwieldy to make much progress here.

A better representation is given by considering the local structure of $\mathcal L$, as alluded to earlier when we introduced current. For the one-dimensional phase space of a bi-infinite monon with transition rates $\rate(i\mapsto i+1)$ and $\rate(i+1\mapsto i)$, we can write the system in the following two convenient ways:
\begin{align*}
  \dot x_i &= -[\rate(i\mapsto i+1)+\rate(i\mapsto i-1)]x_i \\
           &\phantom{{}={}}+ \rate(i-1\mapsto i)\,x_{i-1} + \rate(i+1\mapsto i)\,x_{i+1}, \\
  S(i\mapsto i+1) &= \rate(i\mapsto i+1)\,x_i - \rate(i+1\mapsto i)\,x_{i+1}.
\end{align*}
Note how $\dot x_i \equiv - S(i\mapsto i+1) + S(i-1\mapsto i)$, as expected. Also keep in mind that the expression for $\dot x_i$ needs to be modified to incorporate sources and sinks. These representations are just recurrence relations, and methods to solve these include elimination, induction, and generating functions. To demonstrate, we solve for the MFPT from $\ket{-s}$ to $\ket{0}$.

To set up the problem, we place an absorbing boundary at $\ket{0}$ such that $x_0=0$ and we force the system by placing a negative unit-strength source at $\ket{-s}$ such that $S(-s\mapsto-s+1) - S(-s-1\mapsto-s)=1$. The boundary condition at $-\infty$ is $x_{-\infty}=0$, as is the current. Therefore, we find that $S(-s-k-1\mapsto-s-k)=0$ for $k>0$ and $S(-s+k\mapsto-s+k+1)=1$ for $s>k\ge0$. That is, the current transports density from the source to the sink. Writing out the current equations,
\begin{align*}
  1 = S(-1\mapsto0) &= \rate(-1\mapsto0)\,x_{-1}, \\
  1 = S(-2\mapsto-1) &= \rate(-2\mapsto-1)\,x_{-2} - \rate(-1\mapsto-2)\,x_{-1}, \\
  1 = S(-3\mapsto-2) &= \rate(-3\mapsto-2)\,x_{-3} - \rate(-2\mapsto-3)\,x_{-2}, \\
  &~\,\vdots \\
  0 = S(-s-k-1\mapsto-s-k) &= \rate(-s-k-1\mapsto-s-k)\,x_{-s-k-1} \\&\phantom{{}={}}- \rate(-s-k\mapsto-s-k-1)\,x_{-s-k}, \\
  &~\,\vdots,
\end{align*}
we can find the densities in the vicinity of the boundary;
\begin{align*}
  x_0 &= 0, \\
  x_{-1} &= \frac1{\rate(-1\mapsto0)}, \\
  x_{-2} &= \frac1{\rate(-2\mapsto-1)}\left[1+\frac{\rate(-1\mapsto-2)}{\rate(-1\mapsto0)}\right], \\
  x_{-3} &= \frac1{\rate(-3\mapsto-2)}\left[1+\frac{\rate(-2\mapsto-3)}{\rate(-2\mapsto-1)}\left[1+\frac{\rate(-1\mapsto-2)}{\rate(-1\mapsto0)}\right]\right], \\
  &~\,\vdots.
\end{align*}
More generally, we can write
\begin{align*}
  x_{-n} &= \begin{cases}
    \frac{1}{\rate(-1\mapsto0)}\left(\sum_{k=1}^n\prod_{r=2}^{k}\frac{\rate(-r+1\mapsto-r)}{\rate(-r\mapsto-r+1)}\right), & n \le s, \\
  \frac{1}{\rate(-1\mapsto0)}\left(\sum_{k=1}^s\prod_{r=2}^{k}\frac{\rate(-r+1\mapsto-r)}{\rate(-r\mapsto-r+1)}\right)\left(\prod_{r=s+1}^{n}\frac{\rate(-r+1\mapsto-r)}{\rate(-r\mapsto-r+1)}\right), & n > s.
  \end{cases}
\end{align*}
That is, we have exact expressions for the entire steady state distribution. Adding these up will yield the MFPT, $\tau=\sum_{i=0}^\infty x_i$. This expression is not particularly convenient; we can instead rewrite the recurrence relation as
\begin{align*}
  x_{-n} &= \frac{S(-n\mapsto-n+1)}{\rate(-n\mapsto-n+1)} + \frac{\rate(-n+1\mapsto-n)}{\rate(-n\mapsto-n+1)}x_{-n+1},
\end{align*}
from which we find
\begin{align*}
  \tau &= \sum_{n=1}^{s}\frac{1}{\rate(-n\mapsto-n+1)} + \tau\sum_{n=1}^\infty\frac{\rate(-n+1\mapsto-n)}{\rate(-n\mapsto-n+1)}\frac{x_{-n+1}}{\tau} \\
  &= \left(1-\ev{\frac{\rate(-n+1\mapsto-n)}{\rate(-n\mapsto-n+1)}}\right)^{-1}\sum_{n=1}^{s}\frac{1}{\rate(-n\mapsto-n+1)},
\end{align*}
where the expectation value is taken over the steady state probability distribution. Even if the steady state distribution is unknown, we can nevertheless use this to obtain bounds on $\tau$. In the specific case of homogeneous bias, $\rate(n\mapsto n+1)=p\lambda$ and $\rate(n+1\mapsto n)=q\lambda$, the bounds coincide and thus we can obtain an exact expression without knowledge of the exact distribution (though the distribution also reduces to a reasonably simple form), finding
\begin{align*}
  \tau &= \left(1-\frac qp\right)^{-1}\frac{s}{p\lambda} = \frac{s}{b\lambda},
\end{align*}
as expected.

\para{Generating Functions}

When the values of $\lambda$ take on a particularly nice form, such as constant uniform, generating functions provide a tempting approach. A generating function is used to represent a series as coefficients of a power series. For example, the series $\{0,1,2,3,\ldots\}$ has generating function $t(1-t)^{-2}$ because its series expansion is $t+2t^2+3t^3+4t^4+\cdots$. Using multiple variables, complicated structures such as walks can be encoded in a generating function\footnote{For introductions to generating functions and enumerative combinatorics, see \textcite{gen-fun,enum-comb}. For a review of techniques used to analyse walks in two or more dimensions with generating functions, see \textcite{bm-walks}.}. We will seek a generating function, $W$, whose terms correspond to walks starting at each position $\ket{s}$, and end when they reach $\ket{0}$ for the first time. For convenience, we are walking down towards 0 from positive states, rather than up from negative states. It turns out that the easiest way to construct these walks is in reverse: we start from $\ket{0}$ and add either an `up' step or a `down' step, providing we never go down to $\ket{-1}$. Stated another way, a walk is either the empty walk (with generating function $1$), another walk $w$ followed by an `up' step ($wu$), or another (positive) walk $w_+$ followed by a `down' step ($w_+d$). This way, the walks described are only those right up to the point they cross the origin.

To get meaningful statistical information, we label each step with its probability. In the reversed walk, an `up' step has probability $p$ and a `down' step $q$. This leads to the implicit functional equation
\begin{align*}
  W &= 1 + pxtW + q\bar xt(W-W|_{x=0})
\end{align*}
where $\bar x\equiv 1/x$, $x$ is a variable that records the final position of the walk, and $t$ is a variable that records the number of steps in the walk. To prevent a downward step at $\ket{0}$, we ignore these walks when constructing the next step. Expanding $W$, we can see that it has the terms we expect:
\begin{align*}
  W &= t^0(x^0) + t^1(px^1) + t^2(ppx^2 + pqx^0) + t^3(pppx^3 + ppqx^1 + pqpx^1) + \cdots
\end{align*}
To proceed, we seek a closed form expression. Rewriting, we have $(x-t(px^2+q))W = x - qtW|_{x=0}$. It would appear that we are now stuck, as there is an unknown $W|_{x=0}$ which depends on the complete generating function $W$. Fortunately there is a powerful technique that can be used here, the Kernel method~\cite{gf-kernel}. We require that $W$ have no negative powers of $x$, and therefore the \emph{kernel}, $x-t(px^2+q)$, must be a factor of the right hand side of the equation. Among other things, this means that when the kernel vanishes, so must the right hand side. The kernel has two roots, $X_\pm$, although only one of these turns out to be appropriate (in the sense of not containing negative powers of $t$):
\begin{align*}
  X_\pm &= \frac{1\pm\sqrt{1-4pqt^2}}{2pt}, & 0 &= X_- - qtW|_{x=0}.
\end{align*}
Substituting into our functional equation, we obtain
\begin{align*}
  W &= \frac{1-\bar xX_-}{1-(px+q\bar x)t}.
\end{align*}
In fact, this is not quite the generating function we want. We want the final step to be into an absorbing boundary, so we define the `true' generating function to be $V = pxtW$. A little thought shows that the coefficient of $x^s$ in $V$ is the probability generating function of the first passage time; that is, the coefficient of $t^nx^s$ is the probability that a walk starting at $\ket{s}$ takes $n$ steps to reach $\ket{0}$. In order to find the MFPT, we want to calculate $\sum_{n=0}^\infty n[t^nx^s]V$ where $[t^nx^s]V$ is the coefficient of $t^nx^s$ in V. In fact, we can obtain a generating function encoding this information by taking the $t$-derivative of $V$, exploiting the fact that $\partial_t t^n=nt^{n-1}$. If we then set $t=1$, we sum up all the $t$ coefficients and therefore obtain the generating function for the MFPTs, parametrised by s. That is, $\dot V|_{t=1}$. Evaluating this gives
\begin{align*}
  \dot V|_{t=1} &= \frac1b\frac{x}{(1-x)^2} = \sum_{s=0}^\infty \frac sbx^s,
\end{align*}
i.e.\ the MFPT for $\ket{s}$ is $s/b$ as expected for discrete time. Converting to continuous time recovers $s/b\lambda$.

Unfortunately, these techniques do not extend easily to multiple dimensions. Nevertheless, there is a growing selection of techniques applicable to two-dimensional quadrant walks. See \textcite{bm-walks} for a review. We had partial success in applying these techniques, and our results are summarised in \Cref{app:gf}, but ultimately we were unable to use these to obtain expressions for the MFPT. Instead, we focussed on obtaining lower and upper bounds by making use of the recurrence relations and properties such as detailed balance; these results are presented in \Cref{sec:constrict-discrete}.

\subsection{Techniques for Continuous Phase Space}\label{sec:meth-cont}

\para{Fokker-Planck Equation}

The phase spaces we are interested in are discrete, but for completeness we shall also consider continuous phase spaces. Many of the definitions for discrete phase space generalise naturally such as the phase regions $\mathcal R'$, $\mathcal R$, $\mathcal P$, $\mathcal P'$, $\mathcal I$, and $\mathcal S$. The distribution $W$ is to be interpreted as a density, but otherwise $W$, $G$, $w_n$, $\varrho$, etc.\ generalise as expected. The boundaries of $\mathcal R$ are enforced by a no-flux boundary condition. The hypersurface slicing is less obvious, and will require us to first define the dynamics of the system.

Being stochastic, the dynamics are conventionally described by a Langevin stochastic differential equation. If $\vec\xi(t)$ is the phase coordinate, then we can write $\dv{t}\vec\xi(t)=\vec\eta(\vec\xi;t)$ where $\vec\eta(\vec\xi;t)$ is the noise term, a family of vectors of random variables indexed by the time coordinate. It is standard to use the decomposition $\vec\eta=\vec h+\vec\theta$ where $\vec h=\ev{\vec\eta}$ is a deterministic `forcing' term, and $\vec\theta=\vec\eta-\ev{\vec\eta}$ is a noise term with zero mean. Moreover, we expect the noise term to be uncorrelated in time, i.e.\ our dynamics should have the Markov property. Following the approach of \textcite{risken} in Chapter~3.4, we find we can write $\vec\theta(\vec\xi;t)=\mathbf g(\vec\xi;t)\vec\Gamma(t)$ where $\vec\Gamma$ is a vector of independent unit-variance zero-mean Gaussian variables, and $\mathbf g$ is a noise strength matrix. Using the Kramers-Moyal expansion, we can derive an equation for the evolution of the probability density $W$,
\begin{align*}
  \dot W &= -\vec\nabla\cdot\underbrace{(\vec\mu-\vec\nabla\cdot\mathbf D)W}_{\vec S},
\end{align*}
where $\vec\mu$ is the drift coefficient and $\mathbf D$ the diffusion matrix, both deriving from $\vec h$ and $\mathbf g$. This equation is known as the \emph{Fokker-Planck} equation (FPE).

We can now define hypersurface slicing for continuous phase space, remaining robust against coordinate transforms. Geodesic paths can be found by integrating the drift coefficient, $\dot{\vec x}=\vec\mu(\vec x)$, from some initial coordinate $\vec x_0$. A hypersurface $\mathcal P_s$ is defined by the locus of points which take the same time $t$ to reach $\mathcal S$. The label $s$ is arbitrary as long as it increases monotonically with $t$; typically $s$ will be chosen to correspond to geodesic path lengths in the `physically relevant' coordinate system, e.g.\ $s=b\lambda t$ in the case of uniform constant bias.

To relate the drift coefficient and diffusion matrix to the computational bias, we need to compare the statistical properties of the two systems. For constant bias, this will be straightforward and will result in a constant drift coefficient and diffusion matrix. Where the bias varies, more care is required to ensure the desired properties are replicated in the continuous case. 

We match the statistical properties for a single degree of freedom. In the discrete case, the relevant distribution is given by the difference between two Poisson distributions. One Poisson distribution represents the number of $\oplus$ tokens received, and the other the number of $\ominus$ tokens. This is known as the \emph{Skellam} distribution, and our parameters are $p\lambda t$ and $q\lambda t$, yielding mean $b\lambda t$ and variance $\lambda t$. In the continuous case, the FPE is given by $\dot W=-\mu W'+DW''$ and its exact solution can be obtained by a routine application of Fourier transforms:
\begin{align*}
  \dot{\widetilde W} &= -ik\mu\widetilde W - k^2D\widetilde W \\
  \partial_t \log\widetilde W &= -(k^2D+ik\mu) \\
  W &= \frac1{\sqrt{2\pi}}W_0 \ast \frac1{\sqrt{2\pi}}\int_{\mathbb R}\dd{k} \exp(ikx-(k^2D+ik\mu)t) \\
    &= \frac1{\sqrt{2\pi}}W_0 \ast \frac1{\sqrt{2\pi}}\int_{\mathbb R}\dd{k} \exp(-Dt\left(k+\frac{i(x-\mu t)}{2Dt}\right)^2-\frac{(x-\mu t)^2}{4Dt}) \\
    &= W_0 \ast \frac1{\sqrt{4\pi Dt}}\exp(-\frac{(x-\mu t)^2}{4Dt}).
\end{align*}
That is, the initial distribution $W_0$ is convolved with a Gaussian of mean $\mu t$ and variance $2Dt$. Comparing with the Skellam distribution, we identify $\mu=b\lambda$ and $D=\tfrac12\lambda$.

\section{Constrictive Case}\label{sec:constrict}

Using these techniques, we will be able to obtain a variety of exact, approximate, and numeric results for the constrictive case for general geometries. In practice, we are more interested in truncated simplicial geometries (\Cref{dfn:simplex}) as their construction is more `natural'. The construction is natural because it comprises $d$ mona which evolve independently at all times except for the moment of synchronisation.

\begin{dfn}[Truncated Simplex]\label{dfn:simplex}
  We refer to our primary geometry of interest as a `truncated simplex'. A simplex is a generalisation of the concept of a triangle to arbitrary dimensions. That is, in dimension one it is a line segment, and in dimension three it is a tetrahedron. The pre-synchronisation subspace $\mathcal P$ of \Cref{fig:synch-simple-2} can be considered to be a right simplex in two dimensions, i.e.\ a right triangle, of infinite extent. It is defined by the set $\{(x,y):x\le0\land y\le0\}$ (restricted to $\mathbb Z^2$ for the discrete case or $\mathbb R^2$ for the continuous). The pre-synchronisation subspace $\mathcal P$ of \Cref{fig:synch-wide-1}, meanwhile, is a \emph{truncated} simplex in two dimensions. Its set definition is $\{(x,y):x\le0\land y\le0\land |x+y|\ge w\}$ where $w$ is the side-length of the constriction surface $\mathcal S$. In $d$ dimensions, the set is given by
  \[ \left\{\vec x:\bigwedge_{i=1}^dx_i\le0\land\left|\sum_{i=1}^d x_i\right|\ge w\right\}. \]
\end{dfn}

It will be useful to determine the sizes of hypersurfaces $|\mathcal P_n|$ for truncated simplices. These hypersurfaces are in fact regular $d-1$ simplices, which in $\mathbb R^d$ have hypervolume
\[ \frac{\sqrt{d}}{(d-1)!}\left(\frac{n+w}{\sqrt{2}}\right)^{d-1}. \]
In $\mathbb Z^d$ we can use generating functions. The size of $|\mathcal P_n|$ is equivalent to the number of ways $n+w-1$ can be made from the sum of $d$ natural numbers. This is conveniently given by the coefficient of $x^{n+w-1}$ in $(1-x)^{-d}$, 
\[ {d+n+w-2 \choose n+w-1}. \]

\subsection{Initial Estimates}\label{sec:constrict-initial}

\para{Information Erasure Model}
We shall begin first with some quick initial estimates in order to better understand the qualitative behaviour of our results. Consider the lateral evolution of some initial distribution $\mathcal I$; whilst this initial distribution can in principle be chosen arbitrarily, it will tend to diffuse to fill the entire hypersurface. Moreover, the timescale of this lateral diffusion will be much shorter than that of the approach of the mean position towards the interfacial region $\mathcal S$ in the case of vanishing bias. To clarify, the bias between two adjacent nodes $x$ and $y$ is given by
\[ b(x\mapsto y) = \frac{\rate(x\mapsto y) - \rate(y\mapsto x)}{\rate(x\mapsto y) + \rate(y\mapsto x)}. \]
Therefore, without significant loss of generality, we assume a substantially delocalised initial distribution. This distribution need not be uniform, but it will have an entropy that scales roughly proportional to $\log|\mathcal P_s|$, where $|\mathcal P_s|$ is the size of the initial hypersurface. In order for an interaction to occur, the distribution must pass through the interfacial hypersurface $\mathcal S\equiv \mathcal P_0$ whereupon it will have a distributional entropy proportional to $\log|\mathcal P_0|$. As the geometry is constrictive, $|\mathcal P_0|\le|\mathcal P_s|$ and therefore this process requires an erasure of information. For constant bias $b$, information can be erased~\cite{earley-parsimony-i} at a maximum rate $2b^2\lambda$ where $\lambda$ is the gross transition rate. We can therefore bound the synchronisation time from below by
\[ \frac1{2db^2\lambda} \log\frac{|\mathcal P_s|}{|\mathcal P_0|} \]
which, for a truncated simplex, is
\begin{align*}
    \frac1{2db^2\lambda} \sum_{k=0}^{d-2}\log\left(1+\frac{s}{w+k}\right) 
  = \underbrace{\frac1{4b^2\lambda} \log\left(1+\frac{s}{w}\right)}_{d=2}.
\end{align*}
Note the factor $d$ in the denominator which arises because each monon is able to independently erase information at the given rate.

This neglects the time for the individual mona to reach the synchronisation surface, which is $s/b\lambda$. For $s\lesssim1/b$, the erasure time will dominate and so the MFPT can be approximated as
\[ \frac{s}{b\lambda} + \frac1{4b^2\lambda} \log\left(1+\frac{s}{w}\right), \]
which shows that there is a `penalty' term for synchronisation
When $s\gtrsim1/b$ it is less clear as it is possible that the mona could perform some or all of the erasure along their journey leading to no penalty at all. From a practical point of view, any penalty in this case is negligible as the journey time will be larger, but it is instructive to understand what occurs in this case in order to understand how the system erases information. If the penalty were eliminated, then it would suggest that the lateral distribution of the mona should collimate into a narrow `beam' in anticipation of the synchronisation surface. This is clearly nonsensical, as the tendency is for the mona to diffuse laterally. A simple way to model the interaction is to divide the $d$ mona into 1 `latent' monon and $d-1$ `precocious' mona. The latent monon is identified as the last monon to arrive, and as such the precocious mona can be assumed to have reached a steady state distribution. For uniform constant bias, each monon's steady state distribution is geometric with $\pr(\mathcal P_n)=\frac bp(\frac qp)^n$ and entropy $\log\frac pb+1+\bigOO{\frac bq}$. For the latent monon to pass through the constriction point, each of the precocious mona's distributions must be erased. As these erasures could in principle occur simultaneously, this gives a lower bound for such a penalty as
\[ \frac{d-1}{2db^2\lambda}\log\frac{p}{bw}. \]
Putting the two results together, we see that the lower bound on the penalty should increase logarithmically with distance before reaching a plateau as $s\gtrsim1/b$.

\para{Quasi-Steady State Approximation}
Whilst the above is reasonable from an information theory perspective, it does not provide a mechanism for information erasure. A simple mechanistic model for synchronisation can be given by the Quasi-Steady State Approximation. This approximation consists of two phases; first, the initial distribution $\mathcal I$ evolves and approaches the interfacial surface $\mathcal S$, which is taken to be reflective. We then assume the distribution to have roughly attained its steady state form, at which point the second phase begins. In  the second phase, there is a steady leak of density through $\mathcal S$ per the transition rates of the system which is assumed to not significantly affect the form of the distribution in $\mathcal P$. 

For constant uniform bias, the steady state is given by $\pr(\mathcal P_n) = |\mathcal P_n|(\frac{q}{p})^n / \sum_{k=0}^\infty |\mathcal P_k|(\frac{q}{p})^k$. Evaluating the normalisation constant is non-trivial for arbitrary $|\mathcal P_k|$, but we find approximately that $\pr(\mathcal P_0)\sim b^dw^{d-1}$. The leak rate is $p\lambda\pr(\mathcal P_0)$, and therefore the Quasi-Steady State approximation gives an exponential decay process with half life $\sim b^{-d}w^{1-d}\lambda^{-1}$. This is the approximate penalty term, giving an overall MFPT
\begin{align*}
  \tau &\sim \frac{s}{b\lambda} + \frac{1}{b^dw^{d-1}\lambda}
\end{align*}
and showing that, whilst in principle an order $1/b^2$ penalty term is possible via information erasure, in practice a $d$-dimensional synchronisation will incur a substantially greater penalty for $d>2$ (as $b\ll1$). Fortunately this can be mitigated, either by replacing the synchronisation with a series of $2$-dimensional interactions or by making the information erasure explicit. Explicating the information erasure in such a case can be achieved using the `naive' approach of making the interfacial hypersurface sticky at the cost of erasing at least \SI{1}{\bit} of information (representing the state \texttt{mononIsMoving}).

\subsection{Discrete Phase Spaces}\label{sec:constrict-discrete}

Whilst the preceding approximations are reasonable, they make significant simplifications that neglect much of the specific dynamics of the system. In order to be more confident in their conclusions, we shall proceed by making use of the analytical techniques discussed in \Cref{sec:methodology}. We begin with the discrete case, being that this corresponds to the underlying geometry of the phase spaces of interest.

\subsubsection{Exact results}\label{sec:constrict-exact}

Recall that the MFPT can be obtained as the total phase density at steady state of the modified system, wherein the initial distribution $\mathcal I$ is instead supplied as a constant forcing term. Equivalently, this modified system can be understood as the `teleportation' of density absorbed at $\mathcal S$ back into the system according to $\mathcal I$. This is because, at steady state, the rate of density loss at the absorbing boundary must exactly balance the rate of the forcing term. Observe now that, if $s=1$ (i.e.\ $\mathcal I\subseteq \mathcal P_1$), then the `teleportation' of density from $\mathcal S$ to $\mathcal I$ resembles a reflective boundary. An unforced system with reflecting boundaries and whose underlying Markov chain is reversible has the convenient property that the current is everywhere vanishing. Such a reversible Markov system with vanishing current permits us to readily obtain the steady state using the principle of \emph{detailed balance}: the densities of each pair of states $x$ and $y$ are such that $[x]\rate(x\mapsto y)=[y]\rate(y\mapsto x)$.

Before continuing, we must first clarify the constraint on $\mathcal I$. We begin by introducing a labelling for the states in each hypersurface $\mathcal P_n$ as $\{(n,c) : c\in\mathcal P_n\}$. It is important to note that the steady state densities in the initial hypersurface, $[(s,c)]$, do not generally correspond in a simple way to the initial distribution $[(s,c)]_0\sim\mathcal I$. In the reflective system, the effective forcing term is given simply by the reflection rates,
\begin{align*}
  [(1,c)]_0 &= \sum_{c'\in\mathcal S} [(0,c')]\rate((0,c')\mapsto(1,c)) \\
            &\equiv [(1,c)] \sum_{c'\in\mathcal S}\rate((1,c)\mapsto(0,c')).
\end{align*}
That is, the only permissible initial distribution is prescribed by the steady state distribution of $\mathcal P_1$ and is in proportion to the forward transition rates. As shall be seen shortly, for the example geometry shown in \Cref{fig:synch-wide-1} this yields the initial distribution
\begin{align*}
  [(1,c)]_0 &= \frac1w\begin{cases}
    \frac12, & \text{$c$ on boundary}, \\
    1,       & \text{otherwise},
  \end{cases}
\end{align*}
which, for large $w$, is effectively uniform. Finally, to ensure that the density sum yields the MFPT, we pick the normalisation such that $[\mathcal I]=\sum_{cc'}[(1,c)]\rate((1,c)\mapsto(0,c'))=\sum_{cc'}[(0,c')]\rate((0,c')\mapsto(1,c))=1$. Bear also in mind that the reflecting boundary is an artefact of this representation, with its density in the original system being zero; therefore, in computing the MFPT, we must remember to neglect its contribution.

\para{Truncated Simplices} In addition to being a relatively simple geometry, the truncated simplex systems have constant uniform bias. This property gives rise to a particularly simple steady state. Suppose the forward and backward rates are given by $p\lambda/d$ and $q\lambda/d$ respectively where $p-q=b$ and $d$ is the dimension of the simplex. More precisely, given that each non-boundary state $(n,c)$ is adjacent to $d$ states $(n-1,c'_i)$ in the forward direction and $d$ states $(n+1,c'_i)$ in the backward direction for $i=1\ldots d$, the rates may be written as $\rate((n,c)\mapsto(n-1,c'_i))=p\lambda/d$ and $\rate((n,c)\mapsto(n+1,c'_i))=q\lambda/d$. Using detailed balance, we can deduce therefore that $[(n,c)]=\frac qp[(n-1,c'_i)]$ for \emph{any} $i=1\ldots d$ and hence $[(n-1,c'_i)]=\frac pq[(n,c)]$. As each pair of adjacent hypersurfaces in a truncated simplex forms a connected subgraph, one can show inductively that $[(n,c)]=[(n,c')]$ for all $c,c'\in\mathcal P_n$ (including boundary states). Combining the results thus far, we obtain $[(n+m,c)] = (\frac qp)^m [(n,c')]$. Finally, we find the normalisation as $\forall c.\,1 = [(0,c)] |\mathcal S| q\lambda$ which leads to the complete description of the steady state distribution and hence the MFPT,
\begin{align*}
  [(n,c)] &= \frac{(\frac qp)^n}{q\lambda |\mathcal S|}, &
  \tau &= \sum_{n=1}^\infty \frac{(\frac qp)^n}{q\lambda}\frac{|\mathcal P_n|}{|\mathcal S|},
\end{align*}
where we have neglected the contribution of the reflective boundary density, $[\mathcal S]$ as this is just an artefact of our representation.

Using the expression for the hypersurface sizes, we can compute the MFPT for an arbitrary dimension $d$. The factor $|\mathcal P_n|/|\mathcal S|$ reduces to $\prod_{k=0}^{d-2}(1+\frac n{w+k})$ and therefore the MFPT is given by $\frac1{q\lambda}\sum_{n=1}^\infty(\frac qp)^n\prod_{k=0}^{d-2}(1+\frac n{w+k})$. Enumerating for dimensions $d=1,2,3$, we find the $d$-dimensional MFPTs $\tau_d$,
\begin{equation}\begin{alignedat}{4}
  \tau_1 &= \frac1{q\lambda}\sum_{n=1}^\infty\left(\frac qp\right)^n&&=\frac1{b\lambda}, \\
  \tau_2 &= \frac1{q\lambda}\sum_{n=1}^\infty\left(\frac qp\right)^n\left(1+\frac nw\right)&&=\frac1{b\lambda} + \frac1w\frac p{b^2\lambda}, \\
  \tau_3 &= \frac1{q\lambda}\sum_{n=1}^\infty\left(\frac qp\right)^n\left(1+\frac nw\right)\left(1+\frac n{w+1}\right)&&=\frac1{b\lambda} + \frac{2w+1}{w(w+1)}\frac{p}{b^2\lambda} + \frac1{w(w+1)}\frac{p}{b^3\lambda}.
\end{alignedat}\label{eqn:mfpts-simplex}\end{equation}
Obtaining an expression for general $d$ is non-trivial, but we can see that the leading order term will be $\sim b^{-d}w^{1-d}\lambda^{-1}$ whenever $w\lesssim 1/b$. When $d\lesssim w$, we can also obtain an approximate expression,
\begin{align*}
\tau_d &\approx \frac1{q\lambda}\sum_{n=1}^\infty\left(\frac qp\right)^n\left(1+\frac nw\right)^{d-1} \\
&= \frac1{q\lambda}\sum_{k=0}^{d-1}{d-1\choose k}\left[\left(\frac{t}{w}\pdv{t}\right)^k\frac1{1-t}\right]_{t=\frac qp} \\
&\approx \frac1{q\lambda}\sum_{k=0}^{d-1}{d-1\choose k}\left[\left(-\frac{1}{w}\pdv{s}\right)^k\frac1{s}\right]_{s=2b} \\
&\approx \sum_{k=0}^{d-1}\left(\frac d{2w}\right)^k\frac{1}{b^{k+1}\lambda}.
\end{align*}

\para{General Geometries} Other geometries can be evaluated in the same way as for truncated simplices. We will now generalise the preceding argument as far as possible. First, we use detailed balance to compute the average state density for each hypersurface,
\begin{align*}
  \frac{[\mathcal P_n]}{|\mathcal P_n|} &= \frac1{|\mathcal P_n|}\sum_{c\in\mathcal P_n}[(n-1,c')]\frac{\rate((n-1,c')\mapsto(n,c))}{\rate((n,c)\mapsto(n-1,c'))} \\
  &= \frac{[\mathcal P_{n-1}]}{|\mathcal P_{n-1}|} \underbrace{ \frac1{|\mathcal P_n|}\sum_{c\in\mathcal P_n}\frac{[(n-1,c')]}{[\mathcal P_{n-1}]/|\mathcal P_{n-1}|}\frac{\rate((n-1,c')\mapsto(n,c))}{\rate((n,c)\mapsto(n-1,c'))} }_{\hat t_n},
\end{align*}
where for each $c\in\mathcal P_n$, $c'\in\mathcal P_{n-1}$ is any node adjacent to $c$ (i.e.\ such that the transition rates between them are non-zero). In so doing, we have generalised the factor $t=\frac qp$, used for constant uniform bias, to $\hat t_n$. These densities can then be summed to obtain an expression for the MFPT, 
\begin{align*}
  \tau &= [\mathcal S] \sum_{n=1}^\infty \frac{|\mathcal P_n|}{|\mathcal S|}\prod_{k=1}^n \hat t_k \\
  &= \ev{\sum_{c'}\rate((0,c)\mapsto(1,c'))}_{c\in\mathcal S}^{-1} \sum_{n=1}^\infty \frac{|\mathcal P_n|}{|\mathcal S|}\prod_{k=1}^n \hat t_k,
\end{align*}
where we have used the normalisation condition to solve for $[\mathcal S]$. 

We can now try to extract general scaling behaviour for a broad class of systems. As we are considering constrictive geometries, $|\mathcal P_n|$ is a monotonically increasing function as $n\to\infty$. Therefore, we must have $\prod^n \hat t_k\to0$ as $n\to\infty$ or else the MFPT will diverge (the physical reason being that $\hat t_k>1$ implies negative bias). Moreover, the $\prod^n\hat t_k$ must, in the limit, decrease faster than $|\mathcal P_n|$ grows. Provided that (most) of the $\hat t_k$ are less than 1 this is usually true as it ensures exponential decay whereas we expect $|\mathcal P_n|$ to grow only polynomially. It is reasonable to assume that regions of $\hat t_k>1$ (negative bias) are brief and rare. Given this assumption, the $\prod^n\hat t_k$ can be characterised by a decay length $\ell_n = \min \{m : -\sum_{k=n}^{n+m}\log\hat t_k\ge 1\}$, i.e.\ the distance from $n$ over which it decays by a factor $e$. This decay length is an estimator of the bias, $\ell_n \sim 1/2b$.

Consider the function $n^{p-1} t^n$ for $p,n>0$ and $0<t<1$; the function increases monotonically until its one turning point, before decaying inexorably towards zero. The turning point can be shown to occur at $n=(p-1)\ell$ where $\ell=-1/\log t$; consequently, the integral under the curve is dominated by the range $[0,(p-1+\bigOO{1})\ell)$. The integral can therefore be approximated as $\sim((p-1+\bigOO{1})\ell)^{p}/p\sim p!\ell^{p}$ by assumption that within this range $t^n\sim\bigOO{1}$. Returning to our general MFPT expression, and using the fact that $|\mathcal P_n|$ is monotonic, we can come to the same approximate conclusion. Interpolating quantities to be continuous in $n$, the turning point occurs when $\partial_n\log|\mathcal P_n|=-\log\hat t_n$. Expanding $|\mathcal P_n|$ as a polynomial in $n$ over this range, we can approximate it by its leading order term as $|\mathcal P_n|=an^{p-1}+\bigOO{n^{p-2}}$, and therefore obtain the same range $\sim[0,p\ell)$. Finally then, the MFPT can be approximated to leading order (up to a constant multiplicative factor of order unity) as
\[\tau\sim a\ev{\sum_{c'}\rate((0,c)\mapsto(1,c'))}_{c\in\mathcal S}^{-1}\frac{p!}{2^p}\frac{1}{b^{p}}\]
and we identify $p$ as the effective dimensionality $d$ of the system.

\subsubsection{Bounded Results}

To extend to the case of $s>1$ we proceed inductively. The MFPT from a node $x$ is given by the recurrence relation 
$\tau(x\mapsto\mathcal S) = (\sum_{x'}\rate(x\mapsto x'))^{-1} + \sum_{x'}\rate(x\mapsto x')\tau(x'\mapsto\mathcal S)$
where the $x'$ are adjacent to $x$. To rewrite this inductively, first pick a contiguous hypersurface $\mathcal X$ lying between $\mathcal I$ and $\mathcal S$ such that all paths from $\mathcal I$ to $\mathcal S$ pass through $\mathcal X$ at least once. It can then be seen that $\tau(i\mapsto\mathcal S)$ where $i\in\mathcal I$ can be written as $\tau(i\mapsto \mathcal X) + \sum_{x\in\mathcal X}p_{x|i}\tau(x\mapsto\mathcal S)$ where $\sum_{x\in\mathcal X}p_{x|i}=1$. This itself can be proven inductively by assumption that any node in the region before $\mathcal X$ can be written thus and using the trivial base case of $\tau(x\mapsto\mathcal X)=0$ where $x\in\mathcal X$. We then introduce the distributional quantity $\tau(\mathcal I\mapsto\mathcal S)=\sum_{i\in\mathcal I}\pr(i|\mathcal I)\tau(i\mapsto\mathcal S)$ to write this as $\tau(\mathcal I\mapsto\mathcal S)=\tau(\mathcal I\mapsto\mathcal X_{\mathcal I})+\tau(\mathcal X_{\mathcal I}\mapsto\mathcal S)$ where $\mathcal X_{\mathcal I}$ is the \emph{unique} distribution of states in $\mathcal X$ as they are first reached from $\mathcal I$.

Using this inductive form, we can rewrite the MFPT as the sum of a series of $s=1$ MFPTs,
$\tau(\mathcal I_s \mapsto \mathcal S) = \sum_{k=0}^{s-1}\tau(\mathcal I_{k+1} \mapsto \mathcal I_{k})$
where $\mathcal I_k$ is the unique distribution over $\mathcal P_k$ as first reached from $\mathcal I_{k+1}$. This immediately yields a first approximation to the MFPT for arbitrary $s$ by assuming that $\mathcal I_k$ is similar to the reflective distribution used in \Cref{sec:constrict-exact}. Calculating for the general geometry restricted to uniform constant bias, we find
\begin{equation}\begin{aligned}
  \tau &\approx \sum_{k=1}^s \frac1{q\lambda}\sum_{n=1}^\infty \frac{|\mathcal P_{k+n}|}{|\mathcal P_{k-1}|}\left(\frac{q}{p}\right)^n \\
  &= \frac1{p\lambda}\sum_{n=0}^\infty \left(\frac qp\right)^n \sum_{k=0}^{s-1}\frac{\mathcal P_{k+n+1}}{\mathcal P_k} \\
  &\equiv \frac1{p\lambda}\opn{\mathcal Z}\left\{\sum_{k=0}^{s-1}\frac{\mathcal P_{k+n+1}}{\mathcal P_k}\right\}\!\left[\frac pq\right]
\end{aligned}\label{eqn:mfpt-lb}\end{equation}
where $\mathcal Z$ is the unilateral $\mathcal Z$-transform.

Unfortunately we have been unable to generally determine $\mathcal I_{n-1}$ from $\mathcal I_n$, but we can extract sufficient information to be able to bound the MFPT from above and below. In the reflective approach for $s=1$, recall that the steady state distribution is given by $[(n+m,c)]=(\frac qp)^m[(n,c')]$ for all $c,c'$. Notice also that the transitions to $\mathcal S$ are precisely those that are to be `absorbed', and those from $\mathcal S$ are precisely those to be `teleported'. The consequence is that the distribution on $\mathcal S$ is the unique distribution $\mathcal I_0$ as earlier defined, induced by $\mathcal I=\mathcal I_1$. Consider then the case $s=2$ where the initial distribution $\mathcal I$ is this `reflective' distribution. The MFPT can be decomposed into $\tau(\mathcal I_2\mapsto\mathcal I_1)+\tau(\mathcal I_1\mapsto\mathcal S)$; we know the first term as it is just that given by our $s=1$ calculation, but the second term requires us to know the $s=1$ MFPT for a uniform initial distribution.

The `reflective' distribution is defined by $[(n,c)]\propto\text{number of forward transitions}$; for example, the distribution for the Truncated 2-Simplex is given ratiometrically as $1:2:2:\cdots:2:1$. The uniform $1:1:1:\cdots:1:1$ distribution can thus be considered as a normalised superposition of the reflective distribution and a pure-boundary distribution $1:0:0:\cdots:0:1$. This is useful because the MFPT is linear, and therefore we can apply the principle of linear superposition. Now, for uniform constant bias in a constrictive geometry, the boundary nodes within a given hypersurface $\mathcal P_n$ must have maximal MFPT because the expected transition direction from the boundary nodes is backwards whereas for all other nodes it is forwards. As a result, the MFPT for a uniform distribution is greater than for a reflective distribution. This applies for each subsequently induced hypersurface distribution, as the effective negativity of the boundary transitions causes it to be `sticky' and attract density to itself.

\para{Lower Bound}
Therefore, for an initially reflective distribution, the estimate of the MFPT given by \Cref{eqn:mfpt-lb} is in fact a lower bound. Moreover, it is a lower bound for any distribution lying between the reflective distribution and the limiting distribution induced by the aforementioned boundary attraction phenomenon. 

\para{Upper Bound}
A trivial upper bound is given by the sum of the MFPTs for pure-boundary distributions, but this is needlessly loose. To obtain a tighter upper bound, we appeal to the distributional superposition,
\begin{align*}
  \tau(\mathcal I_n^{\text{refl.}}) &= \tau(\mathcal I_n^{\text{refl.}}\mapsto\mathcal I_{n-1}^{\text{unif.}}) + \alpha\tau(\mathcal I_{n-1}^{\text{refl.}}) + (1-\alpha)\tau(\mathcal I_{m-1}^{\text{bound.}}) \\
  &= \frac1{p\lambda}\sum_{r=0}^\infty\frac{|\mathcal P_{n+r}|}{|\mathcal P_{n-1}|}\left(\frac qp\right)^r + \alpha\tau(\mathcal I_{n-1}^{\text{refl.}}) + (1-\alpha)\tau(\mathcal I_{m-1}^{\text{bound.}}).
\end{align*}
The value of $\alpha$ can be found by taking the ratiometric vector for the reflective distribution, say $1:2:\cdots:2:1$, and then finding the boundary distribution which makes this up to uniform, i.e.\ $1:0:\cdots:0:1$. The uniform distribution would then be $2:2:\cdots:2:2$, and so the proportion of density in the reflective component is $(1+2+\cdots+2+1)/(2+2+\cdots+2+2)$, or the normalised average $\ev{f_x}/\max_x f_x$ where $f_x$ is the number of forward transitions for node $x$.

The boundary MFPT is harder to find, but we can bound it from above. First consider the exact $s=1$ MFPT when $w=1$; in this case, $\mathcal P_1$ is formed entirely from boundary nodes and therefore its MFPT is precisely the boundary MFPT. In fact, this boundary MFPT is an upper bound on all the other boundary MFPTs as it corresponds to the special case of no interior nodes; when interior nodes do exist, they will have lower MFPTs than the boundary and density in the boundary will diffuse into these interior nodes, thus reducing the boundary MFPT. These definitions of $\alpha$ and boundary MFPTs will become clearer with a concrete example.

\para{Truncated Simplices}
Specialising to truncated simplices, we can quantify these lower and upper bounds. The $s=1$ MFPTs have already been obtained earlier for $d=1,2,3$ in \Cref{eqn:mfpts-simplex}, and so the MFPT lower bounds can be obtained thus,
\begin{align*}
  \tau_1 &= \frac{s}{b\lambda}, \\
  \tau_2 &\ge \frac{s}{b\lambda} + \frac{p}{b^2\lambda}\Delta\psi, \\
  \tau_3 &\ge \frac{s}{b\lambda} + \frac{p}{b^2\lambda}\left(2\Delta\psi + \frac1{s+w} - \frac1w\right) + \frac{p}{b^3\lambda}\left(\frac1w-\frac1{s+w}\right),
\end{align*}
where $\Delta\psi=\sum_{k=w}^{w+s-1}\frac1k=\psi(s+w)-\psi(w)\approx\log(1+\frac sw)$ with $\psi$ the digamma function. The factor $\Delta\psi$ closely resembles the information loss in the synchronisation interaction for 2-simplices, and the coefficient of $p/b^2\lambda$ in $\tau_3$ resembles that for 3-simplices, as expected. Notice that these lower bounds do not plateau as $s\to1/b$ as conjectured in \Cref{sec:constrict-initial}; whilst it is in principle possible from an information theoretic perspective to reach this plateau penalty, the passive dynamics of these interactions preclude this possibility. Nevertheless, at the point where the plateau becomes relevant, the penalty becomes insignificant in comparison to the $s/b\lambda$ for $\tau_2$ and therefore this is of little concern. For $\tau_3$ and above, penalties of order $\bigOO{b^{-3}}$ and above are present and remain significant beyond $s/b\lambda$.

\begin{figure}
  \centering
  \includegraphics[width=.8\linewidth]{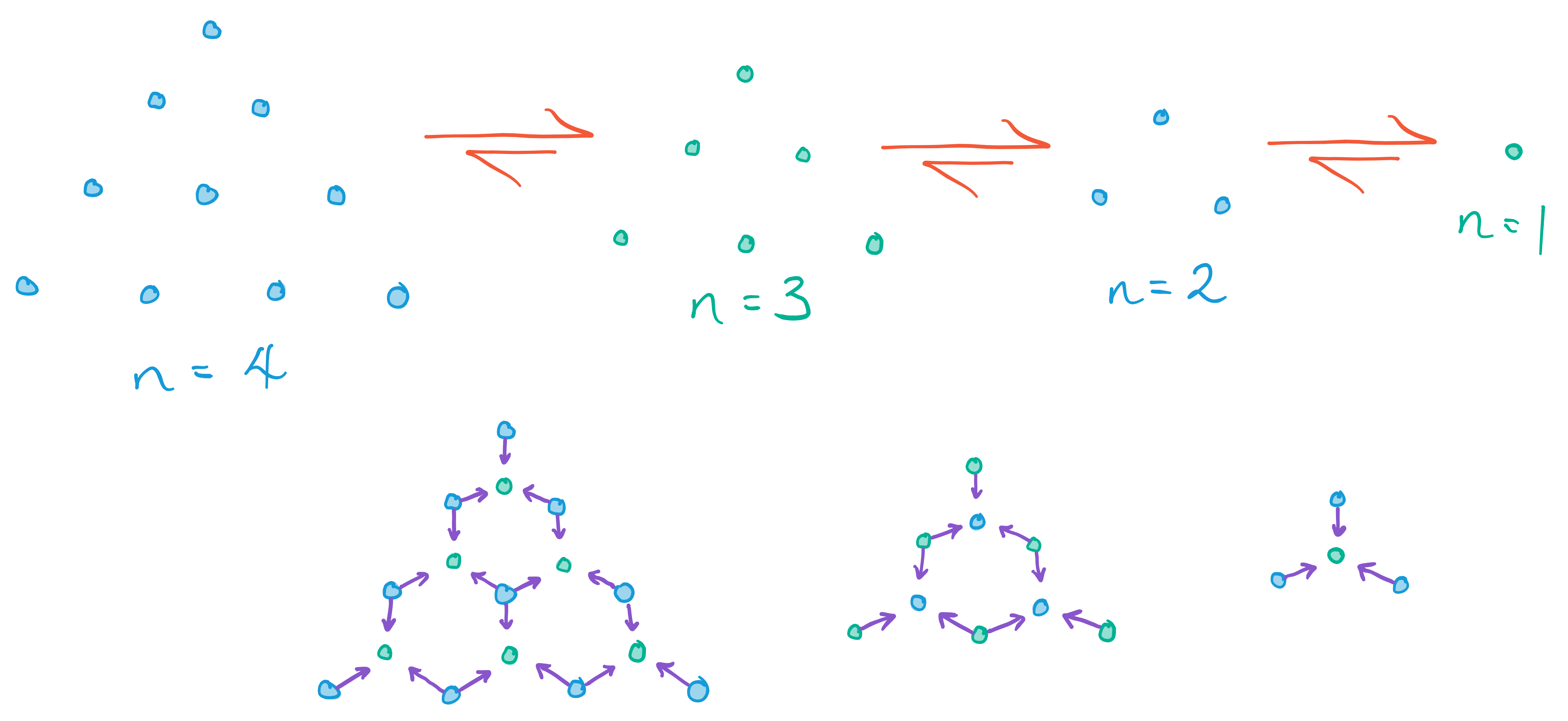}
  \caption{An illustration of a series of hypersurfaces in the $d=3$ simplex to demonstrate the different classes of boundary nodes. The top row shows the hypersurfaces on their own, whilst the bottom row shows adjacent hypersurfaces superimposed and the forward transitions between them.}
  \label{fig:simplex3-boundary}
\end{figure}

For the upper bound, we first obtain the upper bound on the boundary MFPT as $\frac{s}{b\lambda}+\frac{ps}{b^2\lambda}$ for $d=2$ and $\frac{s}{b\lambda}+\frac32\frac{ps}{b^2\lambda}+\frac12\frac{ps}{b^3\lambda}$ for $d=3$. One needs to be particularly careful for $d=3$ and above as there are different `classes' of boundary nodes: for $d=3$ (as shown in \Cref{fig:simplex3-boundary}), there are `vertex' boundary nodes which have only one forward transition, and `edge' boundary nodes which have two forward transitions, whilst interior nodes have three. More generally, in dimension $d$ there are $d-1$ classes of boundary nodes with $1,2,\ldots,d-1$ forward transitions respectively (and interior nodes with $d$ forward transitions). The vertex nodes will have the highest MFPTs and correspond precisely to the $w=s=1$ MFPT, and are therefore what we shall use for MFPT bounds in higher dimensions. The values of $\alpha$ can be found by counting transitions, and are $\alpha_1=1$, $\alpha_2=1-(n+w-1)^{-2}$ and $\alpha_3=1-2(n+w)^{-1}$.

Using these values of the boundary MFPTs and superposition fractions, we can write a recurrence relation for each dimension in $\tau_d(n)$,
\begin{alignat*}{6}
  \tau_1(w+k+1) &= \frac1{b\lambda} +{} &1&\cdot\tau_1(w+k) +{} &0&\cdot\left(\frac{k}{b\lambda}\right), \\
  \tau_2(w+k+1) &\le \frac1{b\lambda} +{} &\frac{w+k-1}{w+k}&\cdot\tau_2(w+k)+{}&\frac1{w+k}&\cdot\left(\frac{k}{b\lambda}+\frac{pk}{b^2\lambda}\right), \\
  \tau_3(w+k+1) &\le \frac1{b\lambda} +{} &\frac{w+k-1}{w+k+1}&\cdot\tau_3(w+k)+{}&\frac2{w+k}&\cdot\left(\frac{k}{b\lambda}+\frac32\frac{pk}{b^2\lambda}+\frac12\frac{pk}{b^3\lambda},
  \right)
\end{alignat*}
with initial conditions
\begin{align*}
  \tau_1(w+1) &= \frac1{b\lambda}, \\
  \tau_2(w+1) &= \frac1{b\lambda} + \frac1w\frac{p}{b^2\lambda}, \\
  \tau_3(w+1) &= \frac1{b\lambda} + \frac{2w+1}{w(w+1)} \frac{p}{b^2\lambda} + \frac1{w(w+1)}\frac{p}{b^3\lambda}.
\end{align*}
These can finally be solved to yield
\begin{align*}
  \tau_1 &= \frac{s}{b\lambda}, \\
  \tau_2 &\le \frac{s}{b\lambda}+\frac{\frac12s(s+1)}{s+w-1}\frac{p}{b^2\lambda}, \\
  \tau_3 &\le \frac{s}{b\lambda}+\frac{\frac12s[2s^2+(3s+1)(w+1)-4s]}{(w+s)(w+s-1)}\frac{p}{b^2\lambda} + \frac{\frac16s(2s^2+3(s-1)(w-1)+4)}{(w+s)(w+s-1)}\frac{p}{b^3\lambda},
\end{align*}
and, in the limit of large $s$, these simplify to
\begin{align*}
  \tau_1 &= \frac{s}{b\lambda}, &
  \tau_2 &\le \frac{s}{b\lambda} + \frac{s+1}{2b^2\lambda}, &
  \tau_3 &\le \frac{s}{b\lambda}+\frac{2s+3w}{2b^2\lambda}+\frac{2s+3w}{6b^3\lambda}.
\end{align*}
In summary, the MFPT penalties for $d=1,2,3$ truncated simplices are bounded above and below by
\begin{align*}
  \Delta\tau_1 &= 0, \\
  \frac{p}{b^2\lambda}\Delta I_2 \le \Delta\tau_2 &\le \frac{p}{b^2\lambda}\frac{\frac12s(s+1)}{s+w-1}, \\
  \frac{p}{b^2\lambda}\Delta I_3 + \frac{p}{b^3\lambda}\frac s{w(s+w)} \le \Delta\tau_3 &\le \frac{p}{b^2\lambda}\frac{\frac12s[2s^2+(3s+1)(w+1)-4s]}{(w+s)(w+s-1)} \\&+ \frac{p}{b^3\lambda}\frac{\frac16s(2s^2+3(s-1)(w-1)+4)}{(w+s)(w+s-1)},
\end{align*}
where $\Delta I_2=\psi(s+w)-\psi(w)$ and $\Delta I_3=2\Delta I_2+\frac1{s+w}-\frac1w$. Expressions for higher dimensions can be obtained using the same approach, but the key point to note is that the terms in each order of $b^{-1}$ are present and significant in both the lower and upper bounds and therefore a synchronisation interaction in dimension $d$ will be subject to a penalty of order $\bigOO{b^{-d}}$, with the coefficient of the $\bigOO{b^{-2}}$ penalty term bounded below by the information loss in the interaction.

\subsubsection{Numerical Simulation}
\label{sec:simulation}

\begin{figure}
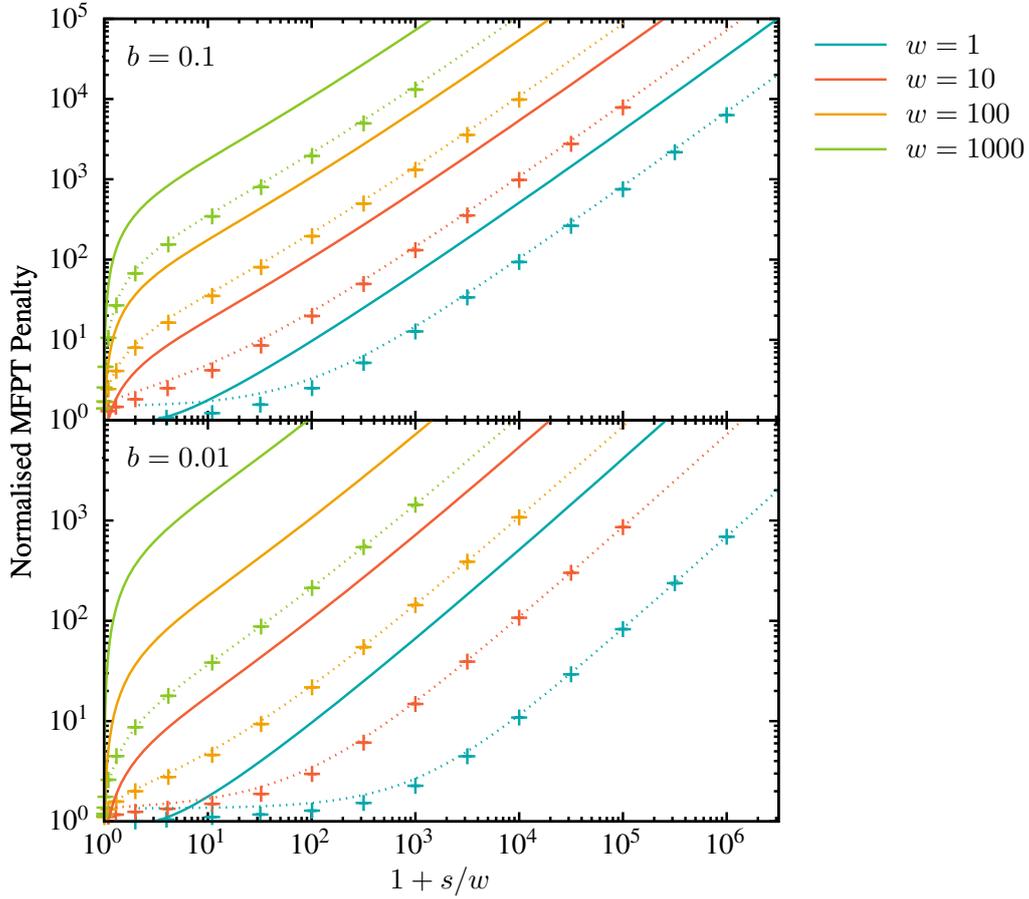

  \captionsetup{singlelinecheck=off,type=figure}
  \centering
  \begin{minipage}[t]{10.5cm}
    \strut\vspace*{-\baselineskip}\newline
    \input{fig-penalties}
  \end{minipage}\begin{minipage}[t]{2.75cm}
    \strut\vspace*{-\baselineskip}\newline
    \\[-3pt]\input{fig-penalties-key}
  \end{minipage}
    
  \caption{Simulation results for the $d=2$ truncated simplex case, as depicted in \Cref{fig:synch-wide-1}, with varying initial hyperplane distances $s$ and constriction widths $w$, and subject to a uniform bias $b\in\{0.1,0.01\}$. The \emph{Normalised MFPT Penalty} corresponds to $(\tau-\frac s{b\lambda})/(\frac p{b^2\lambda}\Delta I_2)$, i.e.\ it is the MFPT penalty divided by the lower bound as computed in \Cref{sec:constrict-discrete}. As a result, the horizontal axis corresponds to the lower bound. The plotted points show simulation data with (negligible) error bars, the solid lines show the (normalised) upper bound, and the dashed lines show the empirical approximation given by \Cref{eqn:mfpt-empirical}.}
  \label{fig:penalties}
\end{figure}

In the course of investigating this synchronisation problem, a suite of tools\footnotemark\ for computing MFPTs was developed. Computing MFPTs in regimes of low bias and unbounded state space is challenging because the probability density tends to diffuse very far from $\mathcal S$. Attempting to compute it in a deterministic way via the explicit steady state distribution has a high spatial complexity as one must make sure not to prematurely truncate the state space, or else risk underestimating the MFPT. A rough estimate gives a spatial complexity of $\bigOO{1/b^2 + (s+w)/b}$, and the time complexity will be the product of this spatial complexity and the number of simulation steps to convergence, which will be at least $\bigOO{1/b}$ to allow particles from the initial distribution to explore the entire state space.
\footnotetext{The suite is available in source-code form at \url{https://github.com/hannah-earley/revcomp-synch-rw-mfpt}. Other tools which may be of interest are available at \url{https://github.com/hannah-earley/revcomp-synch-rw-distribution} and \url{https://github.com/hannah-earley/revcomp-synch-rw-distribution-2}.}

The alternative is to exploit the Markov property of the system in order to use a stochastic Markov-Chain Monte-Carlo simulation approach, or MCMC. In an MCMC approach we essentially apply the stochastic dynamics of the system to an ensemble of instances of the system. In this case, each instance is a single phase particle, whose coordinate represents the joint state of the underlying mona. Therefore, for an ensemble of size $m$, the spatial complexity will be $\bigOO{m}$. Again, the goal is to obtain the steady state distribution of the forced absorbing-boundary system; to ensure conservation of density, we use the teleporting form in which there is a one-to-one correspondence between each particle incident on the absorbing boundary and a particle injected into the initial distribution.

The naive approach is to record the number of time-steps between injection and absorption for each particle. Unfortunately this approach is not robust enough here as the high-diffusion negligible-drift regime leads to substantial variance in the FPT distribution such that a significant proportion of particles in the ensemble will get `lost' in the phase space for an unboundedly long time. Therefore the mean of the particles' injection-absorption times will typically be divergent, or the program will run indefinitely as it waits for each particle to be absorbed. A possible solution is to exclude particles which have not returned after some predefined number of time-steps, but this will lead to an underestimate of the MFPT.

The correct solution, it transpires, is to instead record the absorption current: that is, within some time-step window $\Delta t$, count the number of particles $n$ incident on the absorbing boundary in order to obtain an unbiased estimate of the current $S=n/m\Delta t$ where $m$ is the ensemble size. This process can be repeated as many times as one likes to obtain a series of measurements of the current, from which basic statistics can be used to find an improved estimate of the mean current and its standard error.

There remains, however, a caveat common to all MCMC simulations. Assuming our particles are distributed according to the desired steady state, then their distribution will remain so as they evolve under the stochastic dynamics of the system. The problem is in ensuring that the initial distribution of the system is the steady state, despite not knowing what that is. To address this, one typically initialises the system to an approximation of the steady state and then lets the distribution `burn in' by running the simulation for some number of initial time-steps, after which it is hoped that the approximation will have converged to the steady state. The poorer the quality of the initial approximation to the steady state, the longer this burn-in phase will take. Moreover, the length of the burn-in time can be hard to ascertain and so often one must resort to a heuristic judgement.

Oft forgotten can be the quality of the random number generator used to simulate the stochastic dynamics. In an earlier iteration of this tool, we neglected to consider this and as a result used the standard \texttt{rand} function: a \emph{linear congruential generator} with period $2^{32}$. As the simulation times required to obtain good quality MFPT estimates for many of our parameters exceeded this period, the data obtained was invalid. It can be hard to detect this, and we only realised when a debug trace of the intermediate outputs showed a noticeable and unexpected periodicity. As a result, our MFPT suite now uses the PCG family of PRNGs~\cite{pcg} which have an internal state space size and period of $2^{64}$ whilst also being very fast.

Whilst the dynamics of this system are very simple, some several quadrillion time-steps were required to obtain the data shown in \Cref{fig:penalties} and so a sophisticated toolchain was built around the MCMC routine to improve robustness and automatability. The core program, \texttt{./walk}, is written in \texttt{C+\!+} and makes use of OpenMP for parallelisation. Simulating an ensemble of random walkers is an \emph{embarrassingly parallel} problem, meaning that performance can generally scale linearly with the number of available CPUs due to the minimal amount of inter-thread synchronisation necessary. Unfortunately the \texttt{GCC} and \texttt{LLVM} compilers\footnote{In our experience, \texttt{LLVM}'s optimisation was far better than that of \texttt{GCC} in this case.} were not able to fully optimise the inner loop responsible for executing the stochastic dynamics, and so some hand tuning was necessary to improve the assembly output and bring down the iteration time to $\lesssim\SI{5}{\nano\second}$ on the machines used. To ensure resumable operation and cooperation with cluster scheduling systems such as \texttt{slurm}, the program supports checkpointing (both intermittent and in response to trappable signals). The program is also reasonably modular, making it straightforward to adapt to new random walk systems.

To coordinate simulating the large number of parameters across different topologies of networked systems and to minimise need for manual intervention, a batched job runner was developed in the form of a \texttt{python} script. The job runner has three modes of operation; it can take a description of a set of jobs and generate specific instructions for each job, it can connect to a distributed job queue to request and run these jobs, and it can query the status of all the jobs and job runners (optionally sending updates by email at regular intervals). A key feature is that it can analyse the output of \texttt{./walk} to determine whether the error has converged to a sufficiently small value, whereupon it will move onto the next job. Finally, a number of tools were created to analyse the resulting data, including the ability to infer and inspect the approximate steady state distribution over the phase space.

The results of these simulation tools are shown in \Cref{fig:penalties}. As well as serving as a sanity check on the derived MFPT bounds, we were also able to obtain an estimated empirical equation for the true MFPT as
\begin{align}
  \tau &= \frac{s}{b\lambda}\left(1+p\frac{3+s}{w+s}\right) + \frac43\frac{p}{b^2\lambda}\Delta I_2,\label{eqn:mfpt-empirical}
\end{align}
where the $(3+s)/(w+s)$ term is inspired by the exact MFPT for $b=1$, given by $\tau=s(1+\frac12\frac{3+s}{w+s})$. In particular these tools were very helpful in getting an intuitive feel for the parametric dependence of the MFPT when analytic approaches seemed unyielding.

\subsection{Continuous Phase Spaces}\label{sec:constrict-continuous}

The approach for continuous phase spaces will largely follow that for discrete, except that we shall make use of the Fokker-Planck equation as introduced in \Cref{sec:meth-cont}. In order to solve for the reflective steady state, we shall use the fact that its current vanishes. Recall that the continuous current is given by $\vec S=(\vec\mu-\vec\nabla\cdot\mathbf D)W$ where $W$ is the phase density, $\vec\mu$ the drift coefficient and $\mathbf D$ the diffusion matrix. When $\mathbf D$ is non-singular, this can be rewritten as $-\mathbf D(-\mathbf D^{-1}\vec\mu'+\vec\nabla)W=-\mathbf De^{-\varphi}\vec\nabla e^\varphi W$ where $\vec\mu'=\vec\mu-[\vec\nabla\cdot\mathbf D]$ and $\varphi$ is defined by $\grad\varphi=-\mathbf D^{-1}\vec\mu'$. If the current is identically zero, then it immediately follows that the density is given by $W=Ne^{-\varphi}$ where $N$ is a normalisation constant. This is fairly standard, and a similar result is obtained in Chapter~6 of \textcite{risken}.

To connect this reflective-boundary distribution with the absorbing-boundary distribution, we must find the absorption current. We assume that our reflective-boundary distribution is approximately equivalent to an absorbing-boundary distribution with forcing applied at the $\mathcal P(-\delta x)$ hypersurface, where $x$ is a coordinate orthogonal to the hypersurfaces; in the limit $\delta x\to0$ this becomes exact, in analogy with the discrete case. The true absorbing-boundary distribution has vanishing density at $\mathcal P(0)=\mathcal S$ by definition, and therefore we can recover our desired distribution by introducing an appropriate decay from $\mathcal P(-\delta x)$ to $\mathcal S$. In the limit $\delta x\to0$ any higher order polynomial terms in this decay will vanish, and therefore the decay can be assumed linear. The absorption current is then given by
\begin{align*}
  \vec S(0) &= \vec\mu W|_{x=0} - \sum_i \vec D_i\lim_{\delta x_i\to 0} \frac{W|_{x_i}-W|_{x_i-\delta x_i}}{\delta x_i} = \vec D_x \frac{W(-\delta x)}{\delta x} = \vec D_x \frac{Ne^{-\varphi}}{\delta x}.
\end{align*}
Using the fact that the current is along the $x$ direction, we can find the normalisation constant subject to the constraint that the total absorption current is 1 as $N = \delta x / \int_{\mathcal S}\dd[d-1]{\vec x} D_{xx}e^{-\varphi}$. Therefore we find that the MFPT from $s=-\delta x$ is given by $\delta\tau=\delta x\int_{\mathcal P}\dd[d]{\vec x} e^{-\varphi} / \int_{\mathcal S}\dd[d-1]{\vec x} D_{xx}e^{-\varphi}$, from which we find the lower bound for the MFPT from arbitrary $s$,
\begin{align*}
  \tau(s) &\ge \int_{-s}^0 \dd{x} \frac{\int_{-\infty}^x \dd{x'}\int_{\mathcal P(x')}\dd[d-1]{\vec x} e^{-\varphi}}{\int_{\mathcal P(x)}\dd[d-1]{\vec x} D_{xx}e^{-\varphi}}
    = \int_0^\infty\dd{t}\int_{-s}^0\dd{x}\frac{[\ev{e^{-\varphi}}A]_{x-t}}{[\ev{D_{xx}e^{-\varphi}}A]_x}
\end{align*}
where $\ev{e^{-\varphi}}$ is averaged over the given hypersurface and $A=\int_{\mathcal P(x)}\dd[d-1]{\vec x}$ is the area of that hypersurface. We see that the MFPT for general phase space is easier to express in the continuum limit, and in the particular case of $\varphi$ linear in $x$, i.e.\ $\varphi=\phi-\mu x/D$, this reduces to a Laplace transform:
\begin{align*}
  \tau &\ge \mathcal L_t\!\left\{\int_{-s}^0\dd{x}\frac{[\ev{e^{-\phi}}A]_{x-t}}{[\ev{D_{xx}e^{-\phi}}A]_x}\right\}\!\left(\frac{\mu}{D}\right).
\end{align*}

\para{Truncated Simplices}

For the Truncated Simplex example in $d$ dimensions, the drift coefficient is constant uniform in the $x$ direction and the diffusion matrix is the constant uniform and isotropic $\mathbf D=D\mathbf 1$. Therefore, $\phi=0$ and we have from earlier that
\[A(-n) = \frac{\sqrt{d}}{(d-1)!}\left(\frac{n+w}{\sqrt{2}}\right)^{d-1}.\]
Hence, the MFPT lower bound for arbitrary $d$ may be obtained thus:
\begin{align*}
  \tau &\ge \frac1D\mathcal L_t\!\left\{\int_{-s}^0\dd{x}\left(\frac{w-x+t}{w-x}\right)^{d-1}\right\}\!\left(\frac{\mu}{D}\right) \\
  &\ge \frac1D\mathcal L_t\!\left\{\int_0^s\dd{x}\sum_{k=0}^{d-1}{d+1 \choose k}\left(\frac{t}{w+x}\right)^{k}\right\}\!\left(\frac{\mu}{D}\right) \\
  &\ge \frac1D\sum_{k=0}^{d-1}{d+1 \choose k}\mathcal L_t\{t^k\}\!\left(\frac{\mu}{D}\right)\int_0^s\dd{x}\left(\frac{1}{w+x}\right)^{k} \\
  &\ge \frac{s}{\mu}+(d-1)\frac{D}{\mu^2}\log(1+\frac sw)+\sum_{k=2}^{d-1}\frac{D^k}{\mu^{k+1}}\frac1{k-1}\frac{(d-1)!}{(d-1-k)!}\left[\frac1{w^{k-1}}-\frac1{(w+s)^{k-1}}\right].
\end{align*}

\section{Recessive Case}\label{sec:recessive}

We turn now to the recessive case, for which we shall assume a continuous phase space for simplicity. We shall also restrict our attention to recessive systems whose diffusion matrices are uniform and isotropic, $\mathbf D=D\mathbf 1$, across $\mathcal R=\mathcal P\cup\mathcal P'$ and whose drift vectors are constant uniform in $\mathcal P$; the condition on $\vec\mu$ in $\mathcal P'$ will become apparent later, but for now we shall assume it takes the same value as in $\mathcal P$. Finally, we adopt an orthogonal coordinate system $(u;\vec v)$ where $u$ is chosen such that $\vec\mu=\mu\hat u$. It may appear that $u$ indexes the hypersurfaces, but this is in fact not the case; a coordinate transform that would render $u$ a hypersurface index would apply a shear and therefore make $\mathbf D$ anisotropic.

\begin{figure}
  \captionsetup{singlelinecheck=off,type=figure}
  \centering
  \begin{minipage}[t]{12cm}
    \input{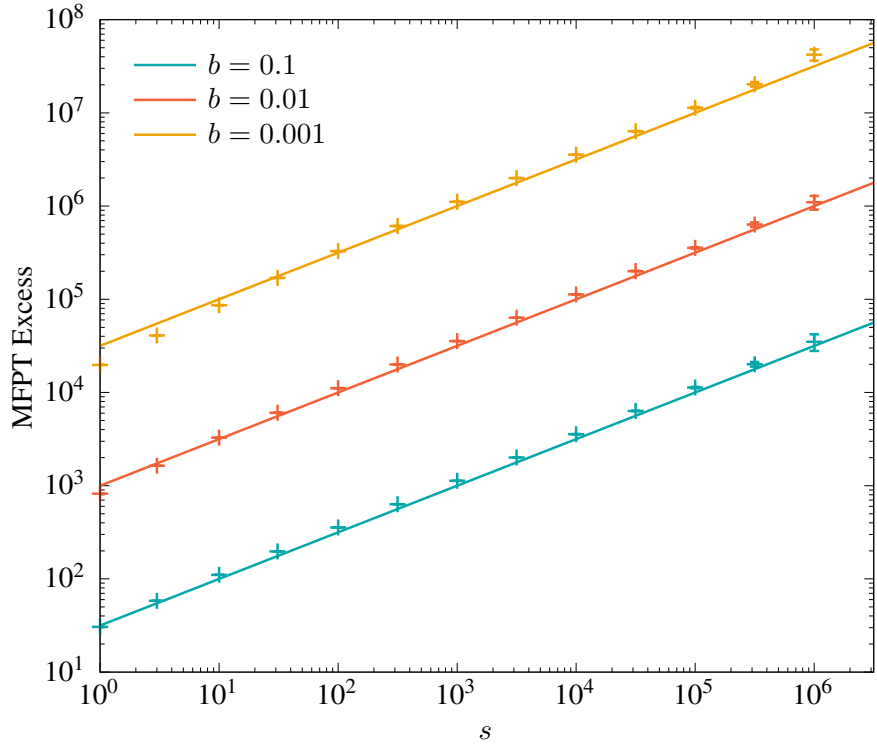}
  \end{minipage}
    
  \caption{Simulation results for the $d=2$ recessive case, as depicted in \Cref{fig:synch-ex-2}, starting on the line $x=y$ and with varying initial distances $s$, and subject to a uniform bias $b\in\{0.1,0.01,0.001\}$. The \emph{MFPT Excess} corresponds to $\tau-\frac{2s}{b\lambda}$ (the coefficient $2$ was chosen empirically). The plotted points show simulation data with error bars, whilst the solid lines show $\frac{\sqrt{s}}{b^{3/2}\lambda}$ hence demonstrating an empirical MFPT of $\frac{2s}{b\lambda}+\frac{\sqrt{s}}{b^{3/2}\lambda}$.}
  \label{fig:gessel}
\end{figure}

\para{MFPT Approach}
Unfortunately the reflective steady state approach used for the constrictive case is inapplicable here as the unbounded transverse width of the space means the steady state is everywhere vanishing. Moreover, ignoring the zero normalisation, the uniform initial distribution is undesirable. For the canonical example shown in \Cref{fig:synch-ex-2}, initial distributions $\mathcal I$ of interest are typically centered on the line $x=y$ (corresponding to the white dashed line in the Figure). Nonetheless, we can apply\footnotemark\ the simulation suite introduced in \Cref{sec:simulation}. The results for an initial delta distribution on the line $x=y$ and a distance $s$ from $\mathcal S$ are shown in \Cref{fig:gessel} and reveal an empirical MFPT of $\tau\approx\frac{2s}{b\lambda}+\frac{\sqrt{s}}{b^{3/2}\lambda}$ and therefore a penalty of order $b^{-3/2}$.
\footnotetext{Our simulation suite is specialised to the case of walks in a single quadrant (with a possibly truncated boundary). Whilst it would not be too difficult to handle more general geometries, this particular three-quadrant example can be easily transformed into a single quadrant walk. First, we identify that the geometry is symmetric in the line $x=y$ and therefore the three quadrants are mapped to one and a half, with a reflective boundary placed at $x=y$. That is, $\mathcal P=\{(-x,y):x\in\mathbb N\land y\in\mathbb Z\land y>-x\}$. Next, we apply a shear $(-x,y)\mapsto(-x,y+x)$ to map the region to the top-left quadrant. The transformed transitions turn out to correspond to those of Gessel walks, a summary of which is presented by \textcite{bm-gessel} along with an investigation of their generating function.}

This penalty would appear to be a significant improvement over that for the constrictive case of $b^{-2}$. Unfortunately this penalty is invalid because in fact the MFPT is not an appropriate quantity to use here. The reason is that the time quantity we are seeking is $\inf\{t:\forall u>0.\ev{n(t+u)}>0\}$ where $n$ is a hypersurface index, and the use of the MFPT is predicated upon the idea that an initial distribution within $\mathcal P'$ has $\ev{n(t)}>0$ for all positive $t$. It is straightforward to see that this property does not apply in the recessive case: Consider a simplified view of \Cref{fig:synch-rec-domain} as just the four quadrant domain $\mathcal R=\mathbb R^2$ such that $\mathcal P=\{(x,y):x<0\wedge y<0\}$ and $\mathcal P'=\{(x,y):x>0\land y>0\}$ and with $\vec\mu=\mu\frac1{\sqrt 2}(\hat x+\hat y)$; the hypersurfaces are given by $\mathcal P(\sqrt{2}n)=\{(n,n+y):y\in\mathbb R_+\}\cup\{(n+x,n):x\in\mathbb R_+\}$, i.e.\ $n(\vec x)=\sqrt 2\min(x,y)$. Now, the evolution of a phase particle may be decomposed into that along the line $x=y$ and perpendicular to this line. Calling these coordinates $u$ and $v$ respectively, i.e.\ $(u,v)=\frac1{\sqrt 2}(x+y,x-y)$ and $n=u-|v|$, we see that the distribution is given by the product of two Gaussians, with probability density
\begin{align*}
  \frac1{\sqrt{4\pi Dt}}\exp(-\frac{(u+s-\mu t)^2}{4Dt}) \cdot \frac1{\sqrt{4\pi Dt}}\exp(-\frac{v^2}{4Dt}).
\end{align*}
Now, the expected hypersurface is given by
\begin{align*}
  \ev{n} &= \ev{u}-\ev{|v|} = -s + \mu t - \sqrt{\frac{4Dt}{\pi}}
\end{align*}
and therefore an initial distribution $\delta(n_0,0)$ will tend to go backwards in $n$-space for a time $t=\frac{D}{\pi\mu^2}$, retreating as far back as $\ev{n}=n_0-\frac{D}{\pi\mu}$ before it begins to advance; at time $t=\frac{2D}{\pi\mu^2}$ it returns to its initial position $\ev{n}=n_0$ whereupon it begins to make net progress. The consequence is that there is a time penalty for all $n_0<+\frac{D}{\pi\mu}$, and hence the appropriate boundary to consider for the MFPT calculation is given by $n=+\frac{D}{\pi\mu}$. 

Of course, in the case of $\vec\mu$ constant uniform across $\mathcal R$ we are able to use the far simpler approach of considering the exact probability distribution, as detailed above. We can also extend easily to arbitrary boundary shape. Unfortunately, as mentioned, $\vec\mu$ may well differ between $\mathcal P$ and $\mathcal P'$. For our canonical example in $(u,v)$ coordinates, 
\begin{align*}
  \vec\mu &= \begin{cases}
    \mu\hat u, & (u,v)\in\mathcal P, \\
    -(\opn{sign}v)\mu\hat v & (u,v)\in\mathcal P'.
  \end{cases}
\end{align*}
That is, in $\mathcal P'$ the drift direction is towards the $u$ axis. Recall also that we have ignored the six-quadrant structure, and so there is in a sense a lot more `space' in $\mathcal P'$. Nevertheless, we will be able to obtain lower and upper bounds.

\para{Probability Distribution Approach}

Ignoring the caveat about $\vec\mu$ for now, we proceed for a general boundary shape $u=f(v)$. For the canonical example, $f(v)=|v|$. Without loss of generality, we pick $f(0)=0$ and we note that a recessive geometry implies $f$ increases monotonically away from $v=0$. We then have $\ev{n}=\ev{u-f(v)}$; decomposing $f$ into its even and odd parts, $f_e(v)=\tfrac12[f(v)+f(-v)]$ and $f_o(v)=\tfrac12[f(v)-f(-v)]$ respectively, we see that $\ev{n}\equiv\ev{u-f_e(v)}$ whenever the initial distribution is even because the transverse dynamics respect evenness. As we are considering the initial distribution $\delta(-s,0)$, we discard the odd part of $f$ without loss of generality. 

If $f$ has a power series expansion about $v=0$, it will take the form $f(v)=\sum_{p=1}^\infty f_{2p}v^{2p}$. Series representations do not exist for many boundary shape functions of interest however, in particular $|v|$ has no such expansion. Consider instead a smooth function $g(v)$ such that $f(v)=g(|v|)$; if $g$ exists, it has series expansion $\sum_{p=1}^\infty f_p v^p$ and hence $f$ will have series expansion $\sum_{p=1}^\infty f_p |v|^p$. We can then compute $\ev{f}$ as $\sum_{p=0}^\infty \alpha_pf_p(2Dt)^{p/2}$ where $\alpha_{2p}=(2p-1)!!$ and $\alpha_{2p+1}=\sqrt{\frac2\pi}(2p)!!$, and $n!!=n(n-2)(n-4)\cdots$ is the double factorial.

For the case $f=f_1|v|$, $\ev{n}=-s+\mu t-2f_1\sqrt{Dt/\pi}$ and we will find a penalty of order $1/\mu^2$ as earlier. If instead $f$ is quadratic, then $\ev{n}=-s+(\mu-2f_2D)t$ and there are two subcases: if $\mu>2f_2D$ then the rate of computation is reduced to $\mu-2f_2D$ within a large vicinity of the synchronisation interaction; if $\mu\le2f_2D$ then computation halts or goes backwards, and the synchronisation never\footnote{It is possible to escape if the phase geometry eventually changes to a permissible form, but the penalty will be substantial.} occurs. As we are considering systems with arbitrarily low bias, and hence $\mu$, this effectively means that quadratic boundaries are untenable. For $f$ cubic and above, $\ev{n}$ briefly increases before decreasing forever. From these, we can deduce that $f$ should ideally be (absolute) linear, though it is admissible to also have a small quadratic component if $\mu$ is bounded from below by $2f_2D$.

We now show how to find lower and upper bounds for $\vec\mu$ per our canonical example, that is $\vec\mu(\mathcal P)=\mu\hat u$ and $\vec\mu(\mathcal P')=-(\opn{sign}v)\mu\hat v$. In both regions, the action of the drift vector alone is to increase $n$. The direction of $\vec\mu$ in $\mathcal P'$ acts to pull the distribution further away from $\mathcal P$ than would be the case if $\vec\mu$ pointed $u$-wards; therefore, picking constant uniform $\vec\mu$ will retard the process of synchronisation and thereby result in an upper bound. Alternatively we can advance the process by picking a uniform superposition of the two, $\vec\mu=\mu(\hat u-(\opn{sign}v)\hat v)$, letting the extra `space' in the $v$ axis apply to $\mathcal P$ too. This extra `space' would appear to complicate matters further, but we can avoid this complication by realising that the action of this $v$-wards drift is simply to increase the value of $n$ and therefore it is equivalent to setting $\vec\mu=2\mu\hat u$. Solving for the bounds finally yields
\begin{align*}
  \tau &\ge \frac{s}{2\mu-2f_2D}+\frac{2f_1^2D}{\pi(2\mu-2f_2D)^2}\left[1+\sqrt{1+\frac{\pi s}{f_1^2D(2\mu-2f_2D)}}\right] \\
  &\le \frac{s}{\mu-2f_2D}+\frac{2f_1^2D}{\pi(\mu-2f_2D)^2}\left[1+\sqrt{1+\frac{\pi s}{f_1^2D(\mu-2f_2D)}}\right],
\end{align*}
which for our canonical example reduces to
\begin{align*}
  \frac{s}{2\mu}+\frac{D}{2\pi\mu^2}\left[1+\sqrt{1+\frac{\pi s}{2\mu D}}\right] \le \tau
  &\le \frac{s}{\mu}+\frac{2D}{\pi\mu^2}\left[1+\sqrt{1+\frac{\pi s}{\mu D}}\right].
\end{align*}
This shows that the penalty is at least of order $\mu^{-2}$ and, for small $s$, is as high as $\mu^{-5/2}$. In other words, the penalty is worse than for the constrictive case. Interestingly however, the penalty is almost exactly the same for higher dimensions if the boundary is replaced by a hypersurface of revolution of $f$; in fact, in dimension $d$ we find
\begin{align*}
  \alpha_p=\frac{(p+d-3)!!}{(d-3)!!}\begin{cases}
    1 & \text{$p,d$ even} \\
    \sqrt{\frac{2}{\pi}} & \text{$p$ odd} \\
    \frac{2}{\pi} & \text{$p$ even, $d$ odd}
  \end{cases}
\end{align*}
but otherwise the form of the synchronisation times remains the same. As a result, using a recessive synchronisation geometry is an alternative way of efficiently synchronising in dimensions $d=3$ and above. Of course, a sequence of $d=2$ constrictive synchronisations will limit the penalty to order $\mu^{-2}$ and so constrictive synchronisation geometries remains the most effective approach.


\section{Conclusion}

The primary result of this paper is that communication between distinct reversible computers in a system with a limiting supply of free energy is very expensive in comparison to independent computation. A single computational step takes time $1/b\lambda$ for gross computational rate $\lambda$ and bias $b=\sqrt{\dot G/k_BT\lambda N}$, where $\dot G$ is the rate of supply of free energy, $N$ is the number of mona, $k_B$ is Boltzmann's constant, and $T$ is the system's average temperature. In contrast, an interaction event such as communication takes at least a time $\sim1/b^2\lambda$. As the bias scales as $\sim\sqrt{A/V}\sim R^{-1/2}$ for a system of convex bounding surface area $A$, internal volume $V$ and radius $R$, communication will tend to `freeze' out as the system gets bigger. This cost is unavoidable as it is due to the implicit erasure of information involved in synchronising divergent computational states; however, if synchronisation is rare then the temporal cost can be reduced. Suppose that the average proportion of mona wishing to communicate at a given time is $\nu\ll1$, then the free energy can be divided into two supplies: $\dot G_{\text{indep.}}$ and $\dot G_{\text{synch.}}$ for independent computation and synchronisation events respectively. A separate set of bias klona can then be introduced for the synchronising mona,
\begin{align*}
  b_{\text{indep.}} &= \sqrt{\frac{\dot G_{\text{indep.}}}{k_BT}\frac{1}{\lambda N}}, &
  b_{\text{synch.}} &= \sqrt{\frac{\dot G_{\text{synch.}}}{k_BT}\frac{1}{\lambda\nu N}}.
\end{align*}
In order for the synchronisation to be viable we therefore require
\begin{align*}
  \nu &\le \frac{\dot G_{\text{synch.}}}{k_BT}\sqrt{\frac{k_BT}{\dot G_{\text{indep.}}}\frac{1}{\lambda N}}.
\end{align*}
We don't want to make independent computation unreasonably slow, and therefore $\dot G_{\text{indep.}}$ will be a substantial fraction of the total supply. To maximise the proportion of mona permitted to synchronise, $\nu$, we take $\dot G_{\text{synch.}}\sim\dot G_{\text{indep.}}$ and hence find $\nu\sim\sqrt{A/V}$. That is, the number of synchronising mona $\nu N$ can be as high as $\sim\sqrt{AV}$. In fact, this subpopulation of mona are also permitted to perform arbitrary operations subject to entropic costs, with the proviso that their individual net transition rates will be $b\lambda\sim R^{-1/2}$. The consequence is that, for a very large reversible computer, any parallel computation should be structured to minimise the need to synchronise the joint state of the system.

Finally, we comment on the applicability of these results to non-Brownian reversible computers in the limit of vanishing free energy density. In the optimal case of ballistic dynamics, there remains an issue if the time of synchronisation cannot be predicted in advance. If so, then as discussed in the introduction there will be dissipation of at least one bit of entropy and a corresponding time penalty $\gtrsim\bigOO{b^{-2}}$. Conversely, if the time at which each computational entity arrives at the synchronisation point can be determined then synchronisation can be achieved without penalty providing that there is no uncertainty, in which case an analysis of \textcite{frank-thesis} shows further gains of reversible computers over irreversible computers. Of course, in practice this is not possible as discussed in Part~I~\cite{earley-parsimony-i}, and so the synchronisation time cannot be effectively predicted. The only way to avoid dissipation in synchronisation is to abolish `true' synchronisation by using a serial architecture (as afforded by the Quantum Zeno architecture defined in Part~I), instead simulating synchronisation using well-established techniques such as pre-emptive multitasking.

\appendix

\section{Acknowledgements}

The author would like to acknowledge the invaluable help and support of his supervisor, Gos Micklem. This work was supported by the Engineering and Physical Sciences Research Council, project reference 1781682.

\section{Constrictive Generating Functions}
\label{app:gf}

In this appendix, we present some limited successes towards generating functions for the MFPT for arbitrary initial distribution. 
%
%
We shall consider the truncated simplex for $d=2$ and $w=1$, as shown in \Cref{fig:synch-simple-2}. To do so, we must construct a function to enumerate all possible walks from each initial position and weight them by their probabilities. A walk starts at some initial position (for convenience we use the upper right quadrant), takes a number of up, down, left and right steps, and terminates at the origin. It is also subject to the constraint that it must not reach the origin until its final step, and if the walker is on a reflective boundary then a forbidden step (left or down, depending on which boundary) is replaced by a null step with the same probability but which leaves its position unchanged.

Constructing a generating function subject to these constraints is easier in reverse. Let the variables of the generating function be $x$, $y$ and $t$, such that a term $\rho x^my^nt^s$ indicates that a walk starting at $(m,n)$ reaches the origin in $s$ steps with probability $\rho\equiv\pr(s|m,n)$. As all walks almost surely reach the origin in a finite time, $\sum_{s=0}^\infty\pr(s|m,n)=1$ and so if $W(x,y;t)=\sum_{mns} \pr(s|m,n)x^my^nt^s$ then $W(x,y;1)=\sum_{mn}x^my^n\equiv\frac1{1-x}\frac1{1-y}$. To construct such a $W$ from walks in reverse, we start from the origin and walk anywhere in the plane for any number of steps. As we shall be handling the origin specially, we start with the last step which must be from either $(0,1)$ or $(1,0)$. The probability that the next step from either of these reaches the origin is $\frac12p$, and therefore these two particular walks are given by the initial term $\frac12pt(x+y)$. Given a walk $w$ of length $s$ starting at $(m,n)$, we can construct a walk of length $s+1$ by prepending it with each of the four possible steps. Ignoring the boundary conditions, this means we can generate a longer walk as $\frac12pt(x+y)w + \frac12qt(\bar x+\bar y)w$ where $\bar x\equiv x^{-1}$. Note that $\frac12ptyw$ corresponds to prepending a downward step with probability $\frac12p$, but we increase the power of $y$ because the power of $y$ is the \emph{starting} position. If $w$ is of the form $\rho x^mt^s$ or $\rho y^nt^s$, then we replace respectively the steps $\frac12qt\bar y$ or $\frac12qt\bar x$ with $\frac12pt$ (this is not a typo; to see that $p$ is correct it is helpful to work through an example walk). Lastly we must ensure walks avoid the origin, or at least that we ignore those walks which prematurely reach the origin. We do so by `trapping' such walks so that they get stuck there by excluding these walks from the $\frac12pt(x+y)w$ transitions. Assembling these, we obtain the functional equation
\begin{align}\begin{aligned}
  W(x,y;t) &= \tfrac12pt(x+y) && \text{(last step)} \\
   &+ \tfrac12qt\bar x[W(x,y;t)-W(0,y;t)] + \tfrac12ptW(0,y;t) && \text{(right step)} \\
   &+ \tfrac12qt\bar y[W(x,y;t)-W(x,0;t)] + \tfrac12ptW(x,0;t) && \text{(up step)} \\
   &+ \tfrac12pt(x+y)[W(x,y;t)-W(0,0;t)], && \text{(left/down step)}
\end{aligned}\label{eqn:functional-constrict-raw}\end{align}
which can be written more usefully in the form
\begin{align}
  \tfrac2p\bar tKW &= (x+y)(1-S) + (1-r\bar y)X(x) + (1-r\bar x)X(y) \label{eqn:functional-constrict} \\
  K &= 1-\tfrac12pt(x+r\bar x+y+r\bar y) \nonumber
\end{align}
where $r=\frac qp$, $S=W(0,0;t)$, $X(x)=W(x,0;t)\equiv W(0,x;t)$, and $K\equiv K(x,y;t)$ is known as the kernel.

\para{Orbital Sum Approach}
We now apply the orbital sum approach as detailed by \textcite{bm-walks}. The group of the kernel is defined to be the set of maps $(x,y)\mapsto(x',y')$ that leave the kernel unchanged, with the group operation taken to be function composition, $\circ$. For this kernel it is generated by the orthogonal involutions $\chi:(x,y)\mapsto(r\bar x,y)$ and $\upsilon:(x,y)\mapsto(x,r\bar y)$, and hence is finite with order 4. Explicitly, the elements of the group are $G=\{1,\chi,\chi\upsilon,\upsilon\}$. We also define the `sign' of each element, $\sigma_\gamma$, as $\sigma_1=\sigma_{\chi\upsilon}=+1$ and $\sigma_\chi=\sigma_\upsilon=-1$. When finite, this group has the interesting property that the `orbital sum' of a function of just $x$ or $y$ under this group vanishes, i.e.\ $\sum_{\gamma\in G}\sigma_\gamma\gamma(f(x))=\sum_{\gamma\in G}\sigma_\gamma\gamma(f(y))=0$.

Looking at our functional equation~\ref{eqn:functional-constrict}, all of the terms on the right hand side are functions of both $x$ and $y$. However, by dividing through by $(1-r\bar x)(1-r\bar y)$ we can make two of these terms functions of only one of $x$ and $y$. We need to be careful however, as generating functions have ring structure but don't generally admit an operation corresponding to division. Moreover, we are dividing through by series in negative powers of $x$ and $y$, and so we need to consider the Laurent series structure. Specifically our series belong to the ring $\mathbb R(x,y)\llb t\rrb$ of formal power series in $t$ whose coefficients are Laurent polynomials in $x$ and $y$, and the series $W$ belongs to the ring $\mathbb R[x,y]\llb t\rrb$ of formal power series in $t$ whose coefficients are polynomials in $x$ and $y$. To perform the division, we observe that the series $-\bar rx(1-\bar rx)^{-1}=-\sum_{n=1}^\infty(\bar rx)^n$ annihilates $(1-r\bar x)$ under multiplication and therefore is an appropriate multiplicative inverse to use.

Multiplying through then and taking the orbital sum, we find
\begin{align*}
  \sum_{\gamma\in G}\sigma_\gamma\gamma\left( \frac{\bar rx}{1-\bar rx}\frac{\bar ry}{1-\bar ry}\tfrac2p\bar tKW \right) &= (1-S) \sum_{\gamma\in G}\sigma_\gamma\gamma\left( (x+y)\frac{\bar rx}{1-\bar rx}\frac{\bar ry}{1-\bar ry} \right).
\end{align*}
Recalling that $K$ is invariant with respect to the group, we can rewrite as
\begin{align*}
  \sum_{\gamma\in G}\sigma_\gamma\gamma\left( \frac{\bar rx}{1-\bar rx}\frac{\bar ry}{1-\bar ry}W \right) &= \tfrac12pt(1-S)\underbrace{\frac{1}{K} \sum_{\gamma\in G}\sigma_\gamma\gamma\left( (x+y)\frac{\bar rx}{1-\bar rx}\frac{\bar ry}{1-\bar ry} \right)}_{R(x,y;t)}.
\end{align*}
To proceed, observe that $\frac{\bar rx}{1-\bar rx}\frac{\bar ry}{1-\bar ry}W$ remains part of the ring $\mathbb R[x,y]\llb t\rrb$. Moreover, every term has a positive power in $x$ and $y$. Therefore, the orbital sum yields the sum of series in the rings $\mathbb R[x,y]\llb t\rrb$, $\mathbb R[\bar x,y]\llb t\rrb$, $\mathbb R[x,\bar y]\llb t\rrb$ and $\mathbb R[\bar x,\bar y]\llb t\rrb$, and so if we extract the positive part consisting of only the terms with positive powers of $x$ and $y$, then we can find an expression for $W$
\begin{align*}
  \frac{\bar rx}{1-\bar rx}\frac{\bar ry}{1-\bar ry}W &= \tfrac12pt(1-S)[x^>y^>]R \\
  W &= \tfrac12pt(1-r\bar x)(1-r\bar y)(1-S)R^>
\end{align*}
where $[x^>y^>]R\equiv R^>$ is the positive part of $R$. There is a limitation here: if $W$ can be written as $f(x,y;t)+g(t)\frac1{1-x}\frac1{1-y}$ then the orbital sum of the $g$ term in this scheme will vanish because $\frac1{1-x}\frac{\bar rx}{1-\bar rx}$ is invariant under the group, as is the $y$ part. Moreover, any other functions which vanish under the orbital sum will also be excluded from our apparent $W$. This means, for example, that evaluating the above expression for $W(x,y;1)$ will yield 0 because $W(x,y;1)$ is in fact $\frac1{1-x}\frac1{1-y}$.

The MFPT starting from $(m,n)$ is given by $\sum_s s\pr(s|m,n)$, and its generating function is $T(x,y)=\sum_{mns} s\pr(s|m,n)x^my^n$. It is easy to show, therefore, that $T(x,y)=\dot W(x,y;1)$ and so we find
\begin{align*}
  T(x,y) &= \tfrac12p(1-r\bar x)(1-r\bar y)(-\dot S(1))R^>(x,y;1),
\end{align*}
assuming that $T$ has no terms which vanish under the orbital sum. It remains to find $S$, and to determine whether or not $T$ has any such terms.

\para{Obstinate Kernel Method}
Another approach is given by the obstinate kernel method, wherein we exploit roots of the kernel. A pair of such roots is given by $(x,Y_\pm(x))$ where $Y_\pm=B\pm\sqrt{B^2-r}$, $B=\frac1{pt}-\tfrac12z$, and $z=x+r\bar x$ is a symmetric function of $x$ and $\chi(x)\equiv r\bar x$. It will be helpful to note the elementary symmetric polynomials of $Y_\pm$ as $\tfrac12(Y_++Y_-)=B$ and $Y_+Y_-=r$. Only one of these roots, $Y_-$, is a power series in $t$; the other includes a term in $t^{-1}$, and so we shall substitute $Y_-$ into the functional equation~\ref{eqn:functional-constrict} to get
\begin{align*}
  0 &= (x+Y_-)(1-S) + (1-Y_+)X(x)+(1-r\bar x)X(Y_-).
\end{align*}
To solve for $X(x)$, and thence $W(x,y)$, we eliminate $X(Y_-)$ by making use of the kernel's group; in particular applying $\chi$ yields a second equation in $X(Y_-)$,
\begin{align*}
  0 &= (r\bar x+Y_-)(1-S) + (1-Y_+)X(r\bar x)+(1-x)X(Y_-),
\end{align*}
from which we find
\begin{align*}
  Q(x)-Q(r\bar x) &= (1-S)(r\bar x-x)\left(\frac{1-Y_--z}{1-Y_+}\right)
\end{align*}
with $Q(x)\equiv(1-x)X(x)$. Extracting the positive part then gives
\begin{align*}
  Q(x)-S &= (1-S)[x^>](r\bar x-x)\left(\frac{1-Y_--z}{1-Y_+}\right).
\end{align*}
From $Q$, $X$ can be obtained and thereby $W$ and $T=\dot W|_{t=1}$. Alternatively, the MFPTs for arbitrary positions can be computed using the recurrence relations and the boundary MFPTs $\dot X|_{t=1}$. In order to do so, we shall need to obtain an expression for $S$. Letting $U(t)=[x^1]X(x;t)$, we see that $[x^1](Q-S)=U-S$ and we compute $[x^0y^0]W$ using \Cref{eqn:functional-constrict-raw} to find $(1-pt)S=qtU$. Therefore we can obtain an expression for $S$,
\begin{align*}
  \frac{S}{1-S} &= \frac{qt}{1-t}[x^1](r\bar x-x)\left(\frac{1-Y_--z}{1-Y_+}\right).
\end{align*}
Using the symmetric variable $x$ helps to evaluate the positive parts of the antisymmetric series $(r\bar x-x)\left(\frac{1-Y_--z}{1-Y_+}\right)$, but we were unable to proceed beyond this point.


\printbibliography

\end{document}